\documentclass[journal=jctcce,manuscript=article]{achemso}
\usepackage{amssymb,amsmath,amsfonts,multicol,multirow,longtable,array,mathpazo}
\usepackage{lscape}
\usepackage[version=3]{mhchem} 
\usepackage{appendix}
\usepackage{soul} 
\usepackage[usenames]{color}
\usepackage{makecell}
\usepackage{enumerate}
\usepackage{xr}
\usepackage[T1]{fontenc} 
\usepackage{booktabs}
\usepackage{tikz}
\usepackage{threeparttable}
\usepackage{graphicx}
\usepackage{subfigure}
\usepackage{colortbl}
\usepackage{framed}
\usepackage{verbatim}
\usepackage{amsbsy}
\usepackage{bm}
\usepackage{bbm}
\usepackage[normalem]{ulem}
\usepackage{float}
\usepackage[linesnumbered,boxed]{algorithm2e}
\usepackage{algorithm2e}
\usepackage{amsmath}

\newcounter{xscheme}

\newfloat{Algorithm}{htbp}{alg}
\floatname{Algorithm}{Algorithm}

\newcounter{exe}[figure]
\newcommand{\iexe}{\refstepcounter{exe}\the\value{exe}:}

\setkeys{acs}{maxauthors = 0} 

\author{Yang Guo}
\affiliation{Qingdao Institute for Theoretical and Computational Sciences, Institute of Frontier and Interdisciplinary Science,
	Shandong University, Qingdao, Shandong 266237, China}
\author{Ning Zhang}
\affiliation{Beijing National Laboratory for Molecular Sciences, Institute of Theoretical and Computational Chemistry,
College of Chemistry and Molecular Engineering, Peking University, Beijing 100871, China}
\author{Yibo Lei}
\affiliation{Key Laboratory of Synthetic and Natural Functional Molecule Chemistry of Ministry of Education, College of Chemistry and Materials Science, Shaanxi Key Laboratory of Physico-Inorganic Chemistry, Northwest University, Xi’an, Shaanxi 710127, China}
\author{Wenjian Liu}\email{liuwj@sdu.edu.cn}
\affiliation{Qingdao Institute for Theoretical and Computational Sciences, Institute of Frontier and Interdisciplinary Science,
Shandong University, Qingdao, Shandong 266237, China}

\title{iCISCF: An Iterative Configuration Interaction-based Multiconfigurational Self-consistent Field Theory for Large Active Spaces}

\setkeys{acs}{articletitle=true}

\begin{document}

\newpage

\begin{abstract}
An iterative configuration interaction (iCI)-based multiconfigurational self-consistent field (SCF) theory, iCISCF, is proposed
to handle systems that require large complete active spaces (CAS). The success of iCISCF stems from three ingredients:
(1) efficient selection of individual configuration state functions spanning the CAS, meanwhile maintaining full spin symmetry;
(2) the use of Jacobi rotation for the optimization of active orbitals, in conjunction with a quasi-Newton algorithm for the core/active-virtual and core-active orbital rotations;
(3) a second-order perturbative treatment of the residual space left over by the selection procedure (i.e., iCISCF(2)).
Just like selected iCI being a very accurate approximation to CASCI, iCISCF(2) is a very accurate approximation to CASSCF.
Several examples that go beyond the capability of CASSCF are taken as showcases to reveal the performances of iCISCF and iCISCF(2).

\end{abstract}

\maketitle

\clearpage
\newpage

\section{Introduction}
It is well known that a multiconfigurational wave function is demanded
even for a qualitative description of a chemical system with a dense set of energetically adjacent frontier molecular orbitals (MO).
The complete active space self-consistent field (CASSCF) theory\cite{FORS1978,CASSCF,WernerRev1987,CASSCFRev1987,MCSCFRev1987,MCSCFRev1998,MCSCFrev2012}
is usually taken as the first step towards an accurate description of such strongly correlated systems. The popularity of CASSCF stems from its operational simplicity:
starting with the partition of the in total $M$ MOs into $M_c$ core (always doubly occupied), $M_a$ active (variably occupied) and $M_v=M-M_c-M_a$ virtual (always zero occupied) orbtials, all configurations are
then generated by distributing the $N_a=N_e-2M_c$ active electrons in the $M_a$ active orbitals in all possible ways, thereby leading
to a subspace full configuration interaction (FCI) problem, usually denoted as CASCI. While the total energy cannot be altered by arbitrary rotations (unitary transformations)
within each of the three subsets of orbitals, the rotations in between do lower the energy further. Therefore, a CASSCF calculation amounts to optimizing the CI and MO coefficients
simultaneously to make the energy stationary. Yet, given so many years of algorithmic developments (see Refs. \citenum{Co-Iter,Werner2020-1,Werner2020-2} for the most recent ones),
the largest CASSCF calculation performed so far involves only 22 active electrons in 22 active orbitals, which was made possible only by massive parallelization\cite{CAS2222}.
The huge computational cost stems of course from the exponential growing of the size of CASCI.
To break the record, some  approximation must be introduced to simplify the CASCI calculation.
As a matter of fact, as long as the approximation can be controlled so as to guarantee the quality of the orbitals, the approximation itself is not an issue at all, simply because
CASSCF is not the end of the calculation but rather the preparation for subsequent treatment of dynamic correlation, which can fully recover
the marginal loss due to the approximate treatment of the CI coefficients.
A large number of approaches have been developed in the last decades to approximate FCI/CASCI (see Ref. \cite{iCIPT2New} for a complete collection and classification of such approaches).
In particular, some of these\cite{v2RDM2008,v2RDM2016,DMRGSCF2008a,DMRGSCF2008b,DMRGSCF2009,DMRGSCF2013,DMRGSCF2014,DMRGSCF2017a,DMRGSCF2017b,FCIQMCSCF2015,FCIQMCSCF2016,HBCISCF2017,HBCISCF2021,iCASSCF2019,ASCISCF,ASCISCF2}
have been adapted to CASSCF, enabling much larger CASSCF calculations. The present work aims to combine the iterative configuration interaction (iCI) approach\cite{iCI} with CASSCF,
leading to iCISCF as a new member of the static-dynamic-static (SDS)\cite{SDS} family of methods (SDSPT2\cite{SDS,SDSPT2}, SDSCI\cite{iCI,SDS}, iCI\cite{iCI}, iCIPT2\cite{iCIPT2,iCIPT2New}, iVI (iterative vector interaction)\cite{iVI,iVI-TDDFT}, $\mathbbm{i}$CAS (imposed automatic selection and localization of active orbitals)\cite{iCAS},
and iOI (iterative orbital interaction)\cite{iOI}).
Among the various unique features\cite{iCIPT2New} of iCI\cite{iCI}, the following ones are particularly relevant to CASSCF:
\begin{enumerate}[(1)]
\item Configuration state functions (CSF) instead of Slater determinants (SD) are taken as the many-electron basis so as to guarantee full spin symmetry, which is of vital importance for describing low-spin states of general open-shell systems.
\item Individual CSFs can be selected with a very efficient algorithm.
\item A particle-hole representation is employed to establish connections between randomly selected CSFs so as to
construct the Hamiltonian matrix very efficiently in a compressed form. In particular, the connections between hole-strings and between particle-strings
can be shared by CSF spaces of different spatial and/or spin symmetries, thereby facilitating the simultaneous calculation of several states of
different spatial and/or spin symmetries with a common set of orthonormal orbitals.
\item The diagonalization of the Hamiltonian matrix is done by iVI\cite{iVI,iVI-TDDFT},
which is able to access directly the roots of a given energy window or specified characters, thereby guiding the SCF iterations to converge to
excited states free of variational collapse  (a point that is not pursued here though).
\end{enumerate}
The iCISCF algorithm is described in Sec. \ref{Theory}, which is followed by pilot applications in Sec. \ref{Results}. The following conventions are to be used throughout:
(1) core, active, virtual, and arbitrary MOs are designated by $\{i,j,k,l,\cdots\}$, $\{t,u,v,w,\cdots\}$, $\{a,b,c,d,\cdots\}$, and $\{o,p,q,r,s,\cdots\}$, respectively;
(2) repeated indices are always summed up.
\section{iCISCF}\label{Theory}
Given a set of MOs that are partitioned into core, active and virtual ones, the selection of individual CSFs in the CASCI space $P$ is
performed iteratively until convergence, with a single parameter $C_{\mathrm{min}}$ for controlling the size of the final, selected space $P_m$.
Specifically, starting with a guess space $P_0$, only those CSFs $\{|I\rangle\in P-P_0\}$ satisfying the ranking criterion $\max_{J\in P_0}|\frac{H_{IJ}C_J}{E^{(0)}-H_{II}}|\ge C_{\mathrm{min}}$ are
put into $P_0$, so as to extend $P_0$ to $P_1$; the Hamiltonian matrix in $P_1$ is then diagonalized\cite{iVI,iVI-TDDFT} to prune away those
CSFs of coefficients smaller in magnitude than $C_{\mathrm{min}}$, so as to reduce $P_1$ to $P_m$ upon convergence (for more details, see Ref. \citenum{iCIPT2New}).
Finally, the Hamiltonian matrix $\mathbf{H}^{(0)}$ in $P_m$, viz.
\begin{eqnarray}
H_{IJ}^{(0)}&=&\langle I|\hat{H}|J\rangle,\quad \forall I, J\in P_m, \label{generalHmat}\\
\hat{H}&=&h_{pq} \hat{E}_{pq} +\frac{1}{2} g_{pq,rs}\hat{e}_{pq,rs},\label{Hoper}\\
\hat{E}_{pq}&=&a_{p\alpha}^\dag a_{q\alpha}+a_{p\beta}^\dag a_{q\beta},\\
\hat{e}_{pq,rs}&=&\{\hat{E}_{pq}\hat{E}_{rs}\}=\{\hat{E}_{rs}\hat{E}_{pq}\}=\hat{E}_{pq}\hat{E}_{rs}-\delta_{qr}\hat{E}_{ps},
\end{eqnarray}
 is diagonalized to yield the normalized iCI wave function
\begin{eqnarray}
|0\rangle =\sum_{|I\rangle \in P_m} |I\rangle C_I^{(0)},\quad \langle I|J\rangle=\delta_{IJ}, \quad \sum_I C_I^2=1, \label{iCI}
\end{eqnarray}
with energy
\begin{eqnarray}
E^{(0)} &=& tr(\mathbf{h}\mathbf{D})+\frac{1}{2}\sum_{rs}tr(\mathbf{J}^{rs}\mathbf{P}^{sr}),\\
D_{pq}&=&\langle 0|\hat{E}_{pq}|0\rangle=D_{qp},\quad \Gamma_{pq,rs}=\langle 0|\hat{e}_{pq,rs}|0\rangle=\Gamma_{qp,sr},\\
J^{rs}_{pq}&=&g_{pq,rs},\quad P^{rs}_{pq}=\frac{1}{2}(\Gamma_{pq,rs}+\Gamma_{qp,rs})=P^{sr}_{pq}=P^{pq}_{rs}.
\end{eqnarray}
Note in passing that the basic coupling coefficients $\langle I |\hat{E}_{pq}|J\rangle$ and $\langle I|\hat{e}_{pq,rs}|J\rangle$ in
the Hamiltonian matrix \eqref{generalHmat} depend only on the relative occupations of the MOs in the CSF pairs $\{|I\rangle, |J\rangle\}$ but not on
the individual MOs. As such, they can be evaluated and reused very efficiently with the tabulated unitary group approach\cite{iCIPT2}.
For later use, we here define $P_s=P-P_m$ as the residual of $P$ left over by the selection, whereas $Q=1-P$ as the complementary space for dynamic correlation.

\begin{figure}[!htp]
	\centering
	\begin{tabular}{c}
		\includegraphics[width=1\textwidth]{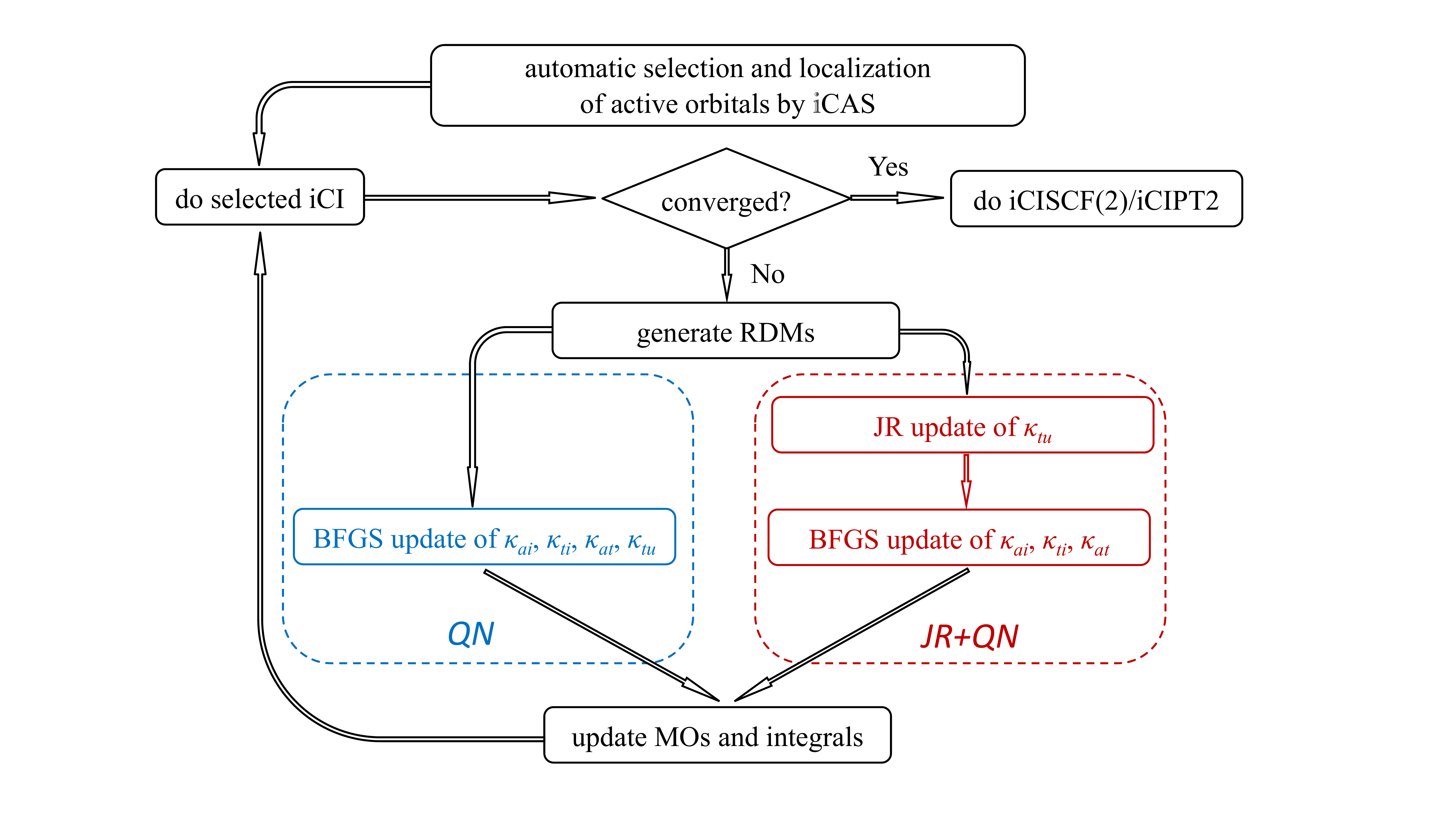}	
	\end{tabular}
	\caption{Flowchart of iCISCF using two different active-active rotation schemes.}
	\label{algorithm}
\end{figure}

To further optimize the orbitals, we parameterize the iCISCF wave function as\cite{ElectronStructure}
\begin{eqnarray}
|\tilde{0}\rangle &=&\sum_{I\in P_m}exp(-\hat{\kappa})|I\rangle C_I,\label{RotPara2} \\
\hat{\kappa}&=&\sum_{pq}\kappa_{pq}\hat{E}_{pq}=\sum_{p>q}\kappa_{pq} \hat{E}_{pq}^-=\frac{1}{2}\sum_{p,q}\kappa_{pq}\hat{E}_{pq}^-,\\
\boldsymbol{\kappa}&=&-\boldsymbol{\kappa}^\dag,\quad
\hat{E}_{pq}^- = \hat{E}_{pq}-\hat{E}_{qp},\quad \frac{\partial\hat{\kappa}}{\partial \kappa_{rs}}=\hat{E}_{rs}^-.
\end{eqnarray}
The skew symmetry of $\boldsymbol{\kappa}$ implies that only $\kappa_{pq} (p>q)$ are independent parameters. Moreover, rotations within the core and within the virtual orbitals are also redundant. Therefore, the $\hat{\kappa}$ operator
actually reads
\begin{eqnarray}
\hat{\kappa}=
\sum_{ai} \kappa_{ai}\hat{E}_{ai}^- + \sum_{ti} \kappa_{ti}\hat{E}_{ti}^- + \sum_{at} \kappa_{at}\hat{E}_{at}^-+ \sum_{t>u} \kappa_{tu}\hat{{E}}_{tu}^-.\label{redKappa}
\end{eqnarray}
The last term in Eq. \eqref{redKappa} vanishes in CASSCF but has to be included in iCISCF due to the truncation of CASCI. It is precisely this term
that renders the orbital optimization of iCISCF more difficult than that of CASSCF.
Minimizing the Lagrangian
\begin{eqnarray}
L[\boldsymbol{\kappa},\mathbf{C}]&=&\langle \tilde{0}|\hat{H}|\tilde{0}\rangle -E(\mathbf{C}^\dag \mathbf{C}-\mathbf{I}) \label{Lagrangian}
\end{eqnarray}
leads to the stationary conditions
\begin{eqnarray}
G^{\mathrm{c}}_I &=&2 \langle I|\hat{H}-E^{(0)}|0\rangle =0,\label{CICond}\\
G^{\mathrm{o}}_{pq}&=&\langle 0|[\hat{E}_{pq}^{-}, \hat{H}]|0\rangle=2\langle 0|[\hat{E}_{pq}, \hat{H}]|0\rangle\label{Orbgrd}\\
&=&2(F-F^\dag)_{pq}=0,\label{OrbCond}\\
\mathbf{F}&=&\mathbf{D}\mathbf{h} + \sum_{rs}\mathbf{P}^{rs}\mathbf{J}^{sr},\label{GenFock}
\end{eqnarray}
where the partial derivatives are evaluated at the expansion point $\boldsymbol{\xi}^{(0)}=(\mathbf{C}^{(0)}, \mathbf{0})^T$.
Eq. \eqref{OrbCond} can be recast\cite{Werner-HF-MCSCF} into a Hartree-Fock-like equation, such that
it can be solved by diagonalization, just like Eq. \eqref{CICond}. However, such first-order orbital optimization method is hardly useful due to very slow convergence.
A more robust method is the second-order Newton-Raphson (QR) scheme, which amounts to expanding the Lagrangian $L$ \eqref{Lagrangian} to second order at the expansion point $\boldsymbol{\xi}^{(0)}$,
\begin{eqnarray}
L^{(2)}[\boldsymbol{\kappa},\mathbf{C}]=E^{(0)}+\boldsymbol{\xi}^{(1)T} \mathbf{G}+\frac{1}{2}\boldsymbol{\xi}^{(1)T} \mathbf{E} \boldsymbol{\xi}^{(1)},
\end{eqnarray}
the stationary condition of which is
\begin{eqnarray}
\mathbf{E}\boldsymbol{\xi}^{(1)}=-\mathbf{G}
\end{eqnarray}
or in block form
\begin{eqnarray}
\begin{pmatrix} \mathbf{E}^{\mathrm{cc}}&\mathbf{E}^{\mathrm{co}}\\
\mathbf{E}^{\mathrm{oc}}&\mathbf{E}^{\mathrm{oo}}\end{pmatrix}\begin{pmatrix}\mathbf{C}^{(1)}\\ \boldsymbol{\kappa}^{(1)}\end{pmatrix}=-\begin{pmatrix} \mathbf{G}^{\mathrm{c}}\\ \mathbf{G}^{\mathrm{o}}\end{pmatrix}, \label{NReq}
\end{eqnarray}
where the second-order partial derivatives (at the expansion point $\boldsymbol{\xi}^{(0)}$) read
\begin{eqnarray}
E^{\mathrm{cc}}_{IJ}&=& \frac{\partial^2 L}{\partial C_I \partial C_J}=2 \langle I|\hat{H}-E^{(0)}|J\rangle,\\
E^{\mathrm{oc}}_{pq,I}&=&\frac{\partial^2 L}{\partial \kappa_{pq} \partial C_I }=2\langle 0|[\hat{E}_{pq}^-, \hat{H}]|I\rangle=E^{\mathrm{co}}_{I,pq},\label{OrbCI}\\
E^{\mathrm{oo}}_{pq,rs}&=&\frac{\partial^2 L}{\partial \kappa_{pq} \partial \kappa_{rs}}
=\frac{1}{2}\langle 0|[\hat{E}_{pq}^-,[\hat{E}_{rs}^-,\hat{H}]]|0\rangle+\frac{1}{2}\langle 0|[\hat{E}_{rs}^-,[\hat{E}_{pq}^-,\hat{H}]]|0\rangle\\
&=&\langle 0|[\hat{E}_{pq}^-,[\hat{E}_{rs}^-,\hat{H}]]|0\rangle+\frac{1}{2}\langle 0|[[\hat{E}_{rs}^-,\hat{E}_{pq}^-],\hat{H}]|0\rangle.\label{Orbhessian}
\end{eqnarray}
The first row of Eq. \eqref{NReq} can be rearranged to
\begin{eqnarray}
2(\mathbf{H}^{(0)}-E^{(0)}\mathbf{I})\mathbf{C}^{(1)}=-\mathbf{G}^{\mathrm{c}}-2\mathbf{H}^{(1)}\mathbf{C}^{(0)},\label{CIresp}
\end{eqnarray}
where
\begin{eqnarray}
H^{(1)}_{IJ}&=&\langle I|\hat{H}_{\kappa}|J\rangle,\label{H1mat}\\
\hat{H}_{\kappa}&=&\frac{1}{2}\kappa_{pq}^{(1)}[\hat{E}_{pq}^-,\hat{H}]=\hat{H}_{\kappa}^\dag \label{Hkoper}\\
&=&[\boldsymbol{\kappa}^{(1)},\mathbf{h}]_{pq}\hat{E}_{pq}+\frac{1}{2}[\boldsymbol{\kappa}^{(1)},\mathbf{J}^{rs}]_{pq}\hat{e}_{pq,rs}+\frac{1}{2}[\boldsymbol{\kappa}^{(1)},\mathbf{J}^{pq}]_{rs}\hat{e}_{rs,pq}\\
&=&[\boldsymbol{\kappa}^{(1)},\mathbf{h}]_{pq}\hat{E}_{pq}+[\boldsymbol{\kappa}^{(1)},\mathbf{J}^{rs}]_{pq}\hat{e}_{pq,rs},\label{Hkappa}
\end{eqnarray}
in which $\hat{H}_{\kappa}$ is a bona fide Hamiltonian operator\cite{ElectronStructure}.
Likewise, the second row of Eq. \eqref{NReq} can be rearranged to
\begin{eqnarray}
(\mathbf{E}^{\mathrm{oo}}\boldsymbol{\kappa}^{(1)})_{pq}&=&-\mathbf{G}^{\mathrm{o}}_{pq}-2\langle 0|[\hat{E}_{pq}^-,\hat{H}|J\rangle C_J^{(1)},\label{Orbresp}\\
(\mathbf{E}^{\mathrm{oo}}\boldsymbol{\kappa}^{(1)})_{pq}&=& \langle 0|[\hat{E}_{pq}^-,\hat{H}_{\kappa}]|0\rangle+[\mathbf{G^{\mathrm{o}}},\boldsymbol{\kappa}^{(1)}]_{pq},\label{OrbOrbsigma}\\
\langle 0|[\hat{E}_{pq}^-,\hat{H}|J\rangle C_J^{(1)}&=&2[\mathbf{D}^{(1)},\mathbf{h}]_{pq}+2\sum_{rs}[\mathbf{P}^{(1)rs},\mathbf{J}^{sr}]_{pq},\label{OrbCIsigma}\\
D^{(1)}_{pq}&=&\frac{1}{2}\langle 0|\hat{E}_{pq}+\hat{E}_{qp}|J\rangle C_J^{(1)},\label{Den1}\\
P^{(1)rs}_{pq}&=&\frac{1}{2} \langle 0|\hat{e}_{pq,rs}+\hat{e}_{qp,rs}|J\rangle C_J^{(1)}.\label{Gamma1}
\end{eqnarray}
Eqs. \eqref{CIresp} and \eqref{Orbresp} can be viewed as response equations for the CI displacements $\mathbf{C}^{(1)}$ and orbital Newton steps $\boldsymbol{\kappa}^{(1)}$, respectively.
The former involves the first-order active space Hamiltonian \eqref{H1mat} due to $\boldsymbol{\kappa}^{(1)}$, whereas the latter involves the first-order reduced density matrices (RDM) \eqref{Den1}/\eqref{Gamma1} due to $\mathbf{C}^{(1)}$.
This particular reformulation\cite{DMRGSCF2017b} is advantageous in that it decouples the orbital optimization from the CI solver implementation in each Newton step, such that any CI solver can readily be used. An even more sophisticated formulation is the Werner-Meyer-Knowles (WMK) approach\cite{SOSCF-Molpro,Werner2020-1,Werner2020-2}, where
the Lagrangian \eqref{Lagrangian} is expanded to second order in $\mathbf{T}=e^{-\boldsymbol{\kappa}}-\mathbf{I}$, thereby containing terms infinite order in $\boldsymbol{\kappa}$.
However, considering that the simultaneous optimization of the CI coefficients and the orbitals suffers from severe linear dependence due to the presence of active-active orbital rotations and that the CI step is rate determining for a large CAS,
 the second term on
the right-hand side of both Eqs. \eqref{CIresp} and \eqref{Orbresp} can be ignored, thereby leading to
\begin{eqnarray}
(\mathbf{H}^{(0)}-E^{(0)}\mathbf{I})\mathbf{C}^{(1)}&=&-\mathbf{G}^{\mathrm{c}},\label{CIresp0}\\
\mathbf{E}^{\mathrm{oo}}\boldsymbol{\kappa}^{(1)}&=&-\mathbf{G}^{\mathrm{o}}. \label{Orbresp0}
\end{eqnarray}
Eq. \eqref{CIresp0} is the usual iterative partial diagonalization of the CI matrix $\mathbf{H}^{(0)}$. For the given RDMs,
Eq. \eqref{Orbresp0} can, in the spirit of quasi-Newton (QR) methods,  be solved iteratively
with the Broyden-Fletcher-Goldfarb-Shanno (BFGS) algorithm\cite{BFGSorg,BFGS}, which amounts to updating iteratively the inverse of the orbital Hessian $\mathbf{E}^{\mathrm{oo}}$ with
the diagonal elements (see Appendix \ref{GHmat}) as the initial guess. While such QN-based CASSCF has been in use for a long time\cite{Gordon1997}, preliminary experimentations
showed that the QN-based iCISCF often does not converge. To handle the troublesome active-active orbital rotations, we adopt the Jacobi rotation (JR) algorithm, which is the most robust first-order optimization method.
In particular, it has been shown\cite{Jacobi} that the convergence pattern of JR-based MCSCF is not much affected when a not fully converged CI vector is used.
In the actual running of iCISCF, the JR algorithm is first employed to optimize the active orbitals until a convergence threshold is fulfilled.
Then, the QN algorithm is invoked to determine the remaining core/active-virtual and core-active rotation parameters. This hybrid algorithm is to be denoted as JR+QN,
to distinguish from the pure QN algorithm.
The flowchart of iCISCF is depicted in Fig. \ref{algorithm}.
If a state-specific Epstein-Nesbet type of second-order perturbation theory (ENPT2) is carried out in $P_s$, the method will be denoted as iCISCF(2). Likewise,
if the ENPT2 is carried out in the joint space $P_s+Q=1-P_m$, the method will be the original iCIPT2\cite{iCIPT2,iCIPT2New}.
Just like that iCIPT2 is very close to FCI\cite{iCIPT2,iCIPT2New,BlindTest}, iCISCF(2) is very close to CASSCF. In principle, the iCISCF(2) energy functional can be minimized
for orbital optimization. However, it has been shown\cite{HBCISCF2017} that this does not change the orbitals discernibly.

\section{Results and discussion}\label{Results}
All calculations were performed with the BDF program package\cite{BDF1,BDF2,BDF3,BDFECC,BDFrev2020} on a computer node
equipped with Intel(R) Xeon(R) Gold 6240 CPUs (in total 36 physical cores) and 128 GB memory.
The convergence threshold for the energy is 10$^{-10}$ $E_{\mathrm{h}}$.
All extrapolations were based on a four-point linear fit of $E_{\mathrm{iCISCF(2)}}$ versus $E^{(2)}$ calculated with different C$_{\mathrm{min}}$,
which is possible because the orbitals with different C$_{\mathrm{min}}$ are virtually identical.

\subsection{Optimization of active orbitals}
Compared with CASSCF, the additional couplings between the active-active orbital rotations and the CI displacements render iCISCF (and other
approximate CASSCF approaches as well) more difficult to converge, especially when they are optimized simultaneously.
To overcome such difficulties, the two sets of parameters have to be decoupled. Even in this case, special care has to be taken of.
For instance, in the adaptive sampling configuration interaction-based CASSCF approach (ASCISCF)\cite{ASCISCF,ASCISCF2},
a sequence of supermacro-, macro-, and micro-iterations is invoked: the selection of a fixed number of SDs is performed
in each supermacroiteration, which consists of a fixed number of macroiterations, where the list of SDs
is kept fixed, such that only the CI coefficients are allowed to respond to changes in the orbitals resulting from the microiterations.
As such, the macro- and micro-iterations together mimic CASSCF (where the list of SDs is fixed by construction).
In contrast, in iCISCF, the selection of individual CSFs is performed in each macroiteration. Since the selection is controlled by a single parameter $C_{\mathrm{min}}$,
both the size and members of the resulting $P_m$ space may change along macroiterations. Therefore, it is of first interest to see
how the optimization of the active orbitals behaves. To this end, the lowest adiabatic singlet and triplet states of hexacene (six linearly fused benzene rings)
are studied with CAS(26,26)/cc-pVDZ\cite{cc-pVDZ} and $D_{2h}$ symmetry. As a matter of fact, identifying the restricted (open-shell) Hartree-Fock (R(O)HF) orbitals as initial guesses for iCISCF is nontrivial for calculations with large active space. However, the pre-CMOs obtained by diagonalizing the molecular Fock matrix projected on the prechosen valence atomic orbitals or the localized pre-CMOs (pre-LMOs)\cite{iCAS}, can be employed to construct initial guess CMOs or LMOs for iCISCF calculations. Thus, in the calculations of hexacene, two different set of initial guess MOs, CMOs and LMOs, constructed by the $\mathbbm{i}$CAS method are used as initial guesses. 

\begin{figure}[!htp]
	\centering
	\begin{tabular}{cc}
		\includegraphics[width=0.5\textwidth]{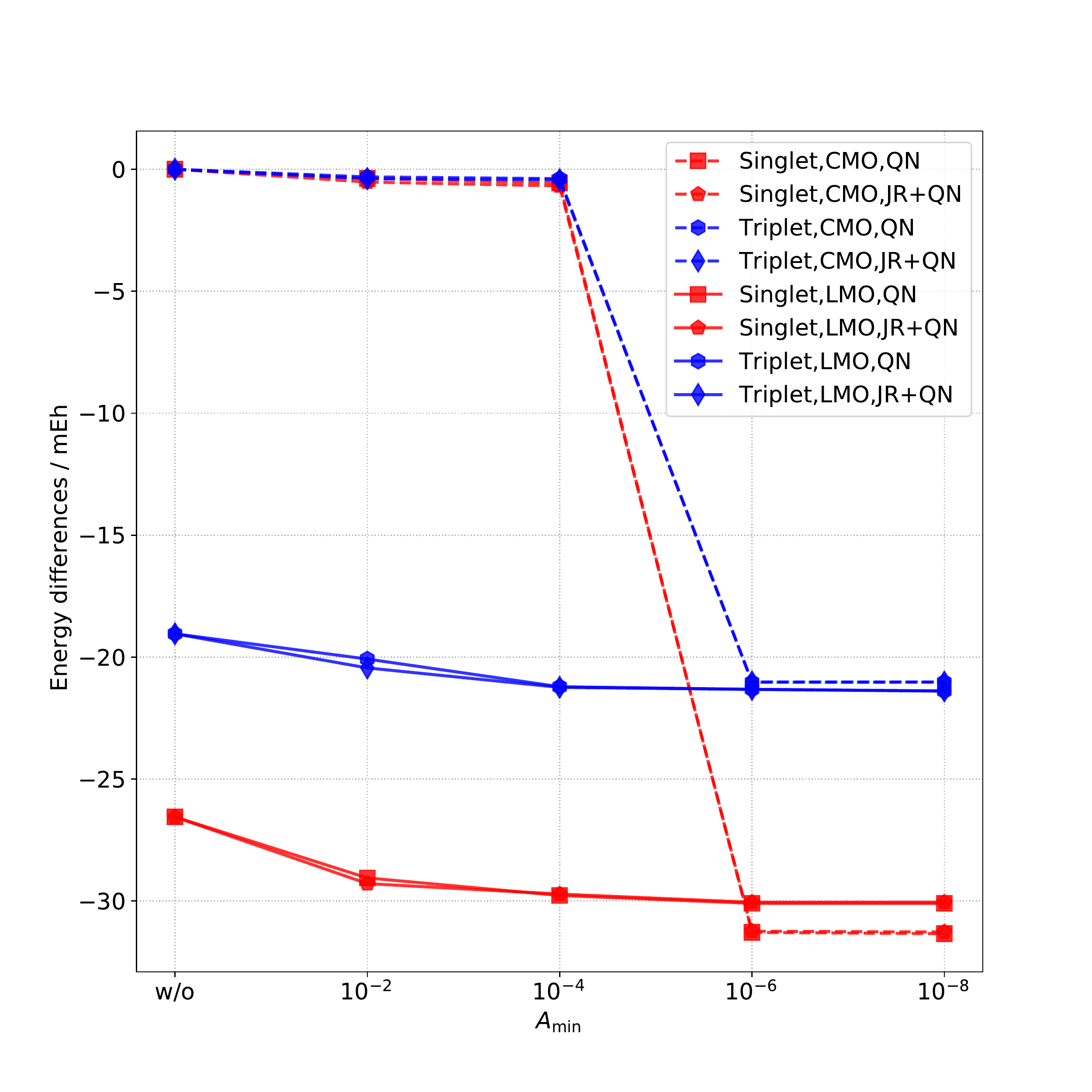}&\includegraphics[width=0.5\textwidth]{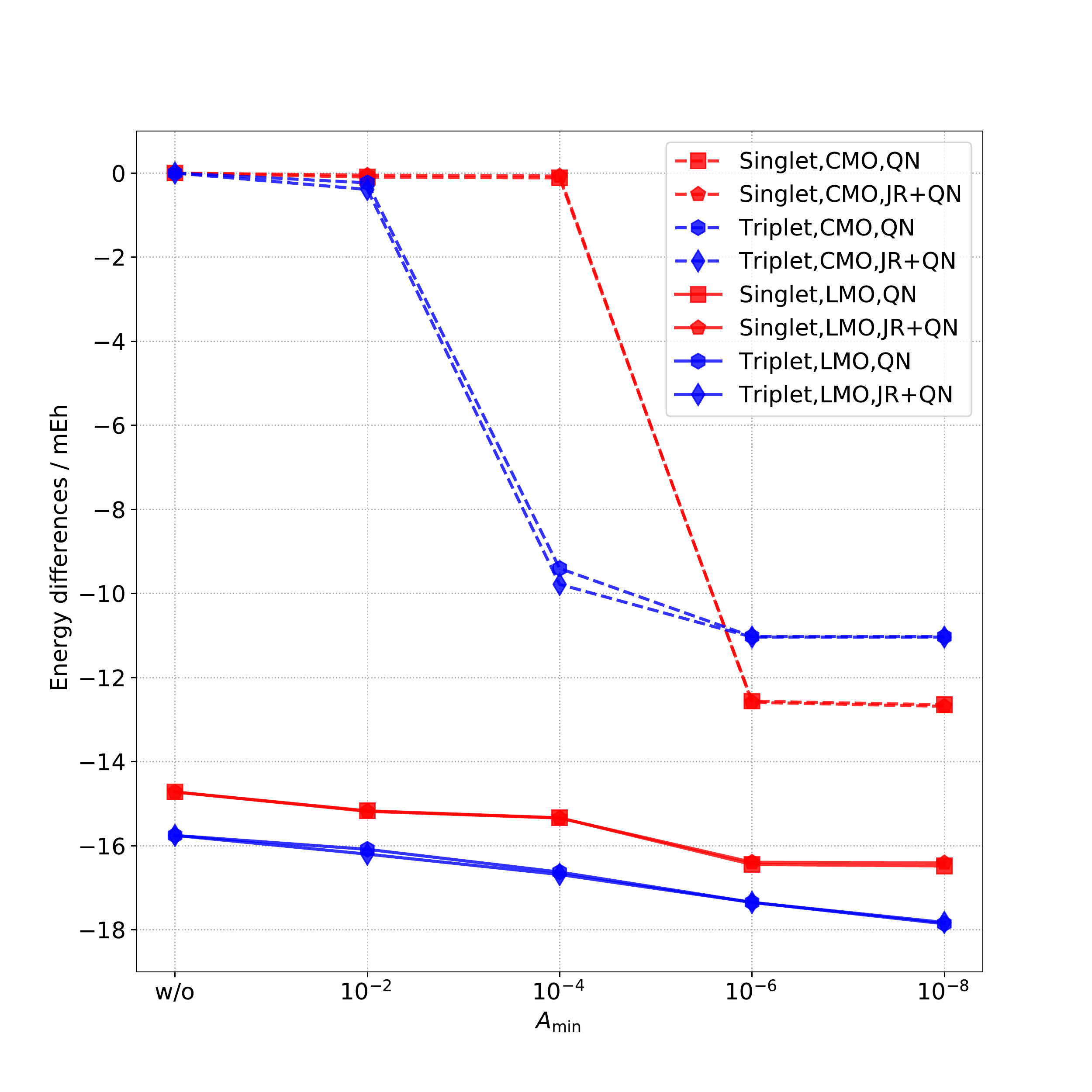}\\
		(a) C$_{\mathrm{min}}=5.0\times10^{-4}$ &(b) C$_{\mathrm{min}}=5.0\times10^{-5}$\\
	\end{tabular}
	\caption{Convergence patterns of the iCISCF calculations of hexacene. The energies without active-active rotations are
taken as zero points. For calculations not converged within 200 macroiterations, the iCISCF energies of the 201$^{st}$ cycle are used.}
	\label{Actopt}
\end{figure}

In both QN and JR+QN algorithms, the optimization of the active orbitals is turned off when the energy difference between two adjacent macroiterations is below the threshold $A_{\mathrm{min}}$. The results with different $A_{\mathrm{min}}$ setting are plotted in Fig. \ref{Actopt},
where the energies without optimizing the active orbitals are taken as the zero points.
It can be seen that both algorithms converge to essentially the same results for
the same $A_{\mathrm{min}}$, and a value of $1.0\times10^{-6}$ $E_{\mathrm{h}}$ is required for $A_{\mathbf{min}}$ to get fully converged results. It is also seen from the Fig. \ref{Actopt} that the initial guess has a grear effect on the convergence of iCISCF results. By using LMOs as initial guess, internal active space optimization is not as important as in the calculations with CMOs. Moreover, the calculations, with different initial MOs, do not always converge to the same results. For example, by setting C$_{\mathrm{min}}=5.0\times10^{-5}$ and $A_{\mathrm{min}}=1.0\times10^{-8}$, the triplet calculations with LMOs guesses deliver lower absolute iCISCF energies than those with CMOs do, which is as large as 6.8 $mE_{\mathrm{h}}$, although the final iCISCF(2) results are close to each other.
   
\begin{figure}[!htp]
	\centering
	\begin{tabular}{c}
		\includegraphics[width=0.8\textwidth]{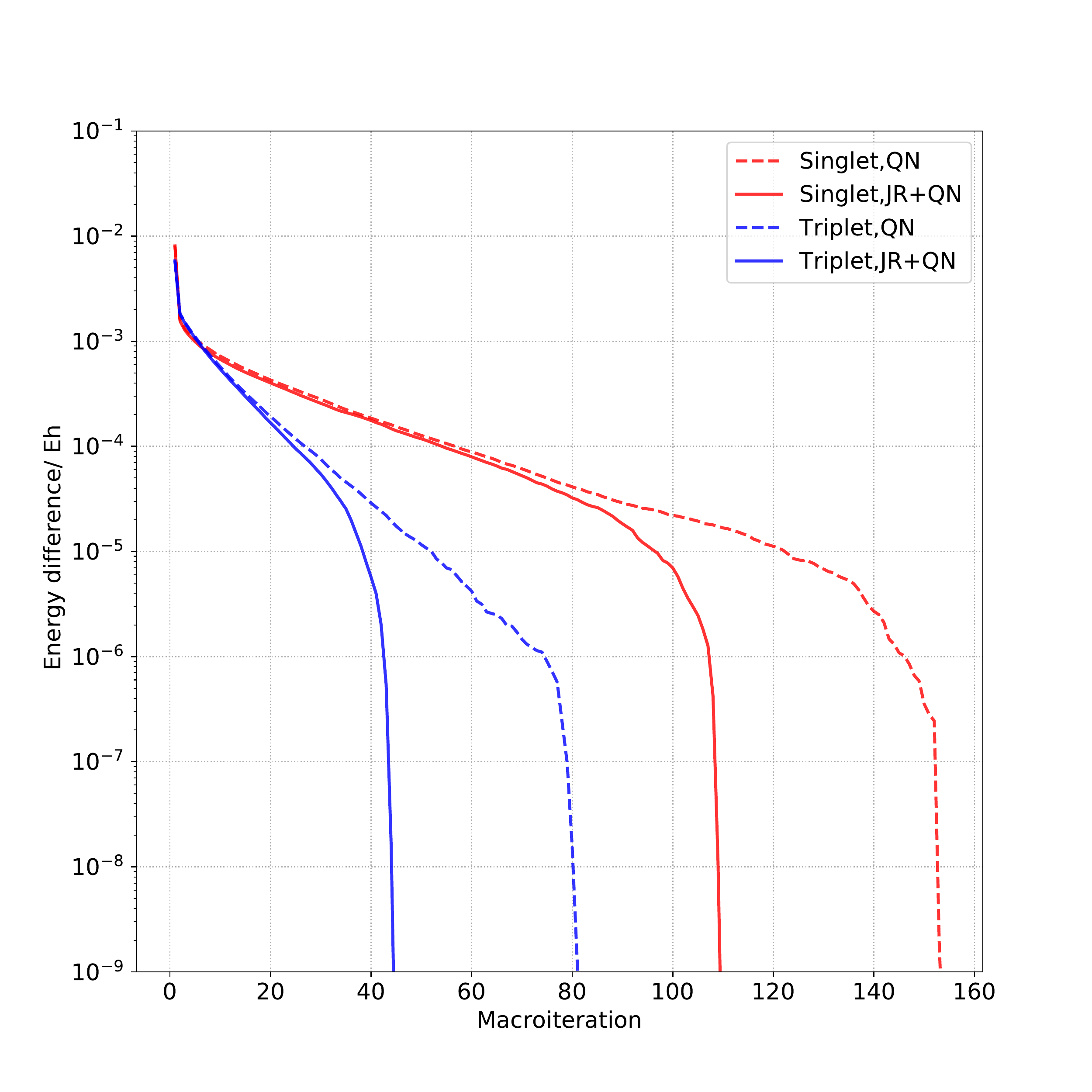}\\
		\\
	\end{tabular}
	\caption{The convergence trends of hexacene calculations using QN and JR+QN algorithms, with C$_{\mathrm{min}}=5.0\times10^{-5}$ and $A_{\mathrm{min}}=1.0\times10^{-8}$ settings, using LMOs as initial guesses.}
	\label{Hexconv}
\end{figure}

The numbers of macro- and microiterations in these calculations are further summarized in Table \ref{Iteration}. The results show that JR+QN does outperform QN in general: QN requires somewhat more microiterations than JR+QN and even fails to converge within 200 macroiterations in two cases. When using a tight $A_{\mathrm{min}}$ threshold, the JR+QN algorithm needs fewer macroiterations to converge for most cases. It can also be seen that using LMOs as initial guesses, the number of macroiteraitons reduces substantially, especially for calculations with the C$_{\mathrm{min}}=5.0\times10^{-4}$ setting. The triplet state calculations converge faster than that of singlet states. However, the calculations for singlet states with C$_{\mathrm{min}}=5.0\times10^{-5}$ and $A_{\mathrm{min}}=1.0\times10^{-8}$ settings still need more than one hundred iterations. The convergence trends of the calculations with these settings optimized by QN and JR+QN algorithms are given in Fig. \ref{Hexconv}.
Within the region from $1.0\times10^{-3}$ to $1.0\times10^{-6} E_{\mathrm{h}}$, all calculations converge slowly. This could be attributed to the coupling between active-active rotation and the CI coefficients.  

In the present work, the JR+QN algorithm is used as the default algorithms for the iCISCF wave function optimization. The LMOs constructed by $\mathbbm{i}$CAS from the pre-LMOs are used as initial guesses for all calculations, if not otherwise specified. By default, $A_{\mathrm{min}}$ is chosen to be 100 times the energy convergence threshold, such that iCISCF has only one parameter,  C$_{\mathrm{min}}$, to play.

\begin{table}[!htp]
	\small
	\centering
	\caption{Number of macroiterations (MA) of the iCISCF calculations of hexacene using the quasi-Newton (QN) and hybrid Jacobi rotation and QN algorithms (JR+QN) with CMOs and LMOs as initial guess, respectively. Averaged number of microiterations (MI) per macroiteration
      are given as well}
	\begin{threeparttable}
		\centering
		\begin{tabular}{cc|cccccccc}\toprule
		   & & \multicolumn{4}{c}{C$_{\mathrm{min}}=5.0\times10^{-4}$ \tnote{a}}
		   &   \multicolumn{4}{c}{C$_{\mathrm{min}}=5.0\times10^{-5}$ \tnote{a}} \\
			&  &  \multicolumn{2}{c}{Singlet} & \multicolumn{2}{c}{Triplet} &  \multicolumn{2}{c}{Singlet} & \multicolumn{2}{c}{Triplet}\\ 
			&  &  MA & MI & MA & MI & MA & MI & MA & MI\\\toprule
&  &     &    &    & \multicolumn{2}{c}{CMOs} & &\\
\multicolumn{2}{c|}{w/o active-active rotations} &6 &3.7&6 &3.8  &7 & 3.3	&8 & 3.0\\
\\
& $A_{\mathrm{min}}$ \tnote{a} \\                                    			
\multirow{4}{*}{ QN}
&  $1.0\times10^{-2}$ &n.c.\tnote{c} &2.2	&7 &5.0 &6 & 6.2	&8 & 5.1          \\
&  $1.0\times10^{-4}$ &	8 &6.8	&7 &6.0           &7 & 7.5	&51 & 14.8        \\
&  $1.0\times10^{-6}$ &128& 11.1	& 133 &12.0     &174 & 12.4	  &78 & 14.5    \\
&  $1.0\times10^{-8}$ &168 &11.1	&147& 12.0      &n.c.\tnote{c} & 12.3	&80 & 14.6     \\
\\                                    			
\multirow{4}{*}{JR+QN}
&  $1.0\times10^{-2}$ &8   &3.6	& 6   &4.0 &7 & 3.6	   &7 & 3.7   \\
&  $1.0\times10^{-4}$ &9   &3.6	& 8   &3.5 &8 & 3.4	   &51 & 3.0  \\
&  $1.0\times10^{-6}$ &125 &3.4&	123 &3.1 &173 & 2.3	 &72 & 2.7\\
&  $1.0\times10^{-8}$ &147 &3.4&	128 &3.1 &n.c.\tnote{c} & 2.3	&78 & 2.7\\
\\
&  &     &    &    & \multicolumn{2}{c}{LMOs} & &\\
\multicolumn{2}{c|}{w/o active-active rotations} &6&5.3	&6 & 5.5	&6 & 6.0	&12 & 3.3\\
\\
& $A_{\mathrm{min}}$ \tnote{a} \\                                    			
\multirow{4}{*}{ QN}
&  $1.0\times10^{-2}$ &5  & 8.2	&6  & 7.5	  &11 & 4.3	&7 & 6.1 \\
&  $1.0\times10^{-4}$ &9  & 12.0	&11 & 10.1	&11 & 7.5	&14 & 9.0\\
&  $1.0\times10^{-6}$ &22 & 10.2	&25 & 11.4	&87 & 8.2	&50 & 8.2\\
&  $1.0\times10^{-8}$ &28 & 8.8	&34 & 10.4	&154 & 5.7	&82 & 5.5\\
\\                                    			
\multirow{4}{*}{JR+QN}
&  $1.0\times10^{-2}$ &6  & 6.2	&6 & 6.0	&8 & 5.0	&7 & 5.3 \\
&  $1.0\times10^{-4}$ &11 & 4.9	&11 & 5.3	&13 & 4.1	&14 & 4.1\\
&  $1.0\times10^{-6}$ &19 & 3.9	&13 & 5.2	&89 & 1.9	&46 & 2.5\\
&  $1.0\times10^{-8}$ &21 & 3.6	&36 & 3.8	&112 & 1.7	&46 & 2.5\\
\midrule
		\end{tabular}
		\begin{tablenotes}
    \item[a]$C_{\mathrm{min}}$: threshold for terminating the selection of individual configuration state functions.
    \item[b]$A_{\mathrm{min}}$: threshold for terminating the active-active orbital rotations.
	\item[c]n.c.: not converged within 200 macroiterations.
    \end{tablenotes}
	\end{threeparttable}\label{Iteration}
\end{table}

\subsection{Adiabatic singlet-triplet gaps of Polyacenes}

The singlet-triplet (S-T) gaps of polyacenes have been studied by several groups using various methods tailored
for large active spaces\cite{DMRG_Radical,v2RDM2016,ACI2018,ASCISCF2}.
To compare with such calculations, the adiabatic S-T gaps are recalculated with iCISCF and iCISCF(2) using the cc-pVDZ basis.
The geometries of naphthalene and C$_{4n+2}$H$_{2n+4}$ ($n$=3-10) are taken from Refs. \citenum{ACI2018} and \citenum{ASCISCF2}, respectively.
The distributions of the orbitals in the irreducible representations (irrep) of $D_{2h}$
are given in Table \ref{IrrepAcene}.
Interestingly, the increments of inactive orbitals in $B_{1g}$ and $B_{2u}$ are different for even and old n-acenes.
As a result, the HOMO and LUMO of even and old n-acenes have different symmetries.
The calculated energies of hexacene are given in Table \ref{Hexacene}. To make a close comparison with ASCISCF running with 10$^6$ SDs\cite{ASCISCF2},
we also report the interpolated iCISCF and iCISCF(2) energies. It can be seen that iCISCF produces slightly lower absolute energies than ASCISCF does for both singlet and triplet state employing the same $P_m$ size. The deviation between the two methods at the variational level is about 3.4 kcal/mol for $^{1}A_{g}$ state and 1.74 kcal/mol for $^{3}B_{u}$ state.
Presumably, this small discrepancy stems from spin contamination in ASCISCF, which is partly removed by
the PT2 correction, thereby leading to a closer agreement between ASCISCF(2) and iCISCF(2) for the S-T gap, within 1.0 kcal/mol. The extrapolated S-T gaps by iCISCF(2) and ASCISCF(2) agree within 1.0 kcal/mol as well.

\begin{table}[!htp]
	\small
	\centering
	\caption{ Distributions of core and active orbitals of n-acenes in irreducible representations of $D_{2h}$}
	\begin{threeparttable}
		\centering
		\begin{tabular}{ccc|cccccccc}\toprule			
			\multicolumn{1}{c}{$n$}
			&\multicolumn{1}{c}{active space}			
			&\multicolumn{1}{c}{}
			&\multicolumn{1}{c}{$A_{g}$}
			&\multicolumn{1}{c}{$B_{3g}$}
			&\multicolumn{1}{c}{$B_{2g}$}
			&\multicolumn{1}{c}{$B_{1g}$}
			&\multicolumn{1}{c}{$A_{u}$}
			&\multicolumn{1}{c}{$B_{3u}$}
			&\multicolumn{1}{c}{$B_{2u}$}
			&\multicolumn{1}{c}{$B_{1u}$}\\\toprule			                                    			
			\multirow{2}{*}{5} & \multirow{2}{*}{(22, 22)}
			& core	&18	&0	&0	&13	&0	&15 &16	&0  \\
			& &active	  &0	&5	&6	&0	&5	&0	&0	&6  \\
			\multirow{2}{*}{6} & \multirow{2}{*}{(26, 26)}
			& core	&21	&0	&0	&16	&0	&17 &19	&0  \\
			& &active	  &0	&6	&7	&0	&6	&0	&0	&7  \\
			\multirow{2}{*}{7} & \multirow{2}{*}{(30, 30)}
			& core	&24	&0	&0	&18	&0	&20 &22	&0  \\
			& &active	  &0	&7	&8	&0	&7	&0	&0	&8\\\midrule
		\end{tabular}
	\end{threeparttable}\label{IrrepAcene}
\end{table}

\begin{table}[!htp]
	\footnotesize
	\centering
	\caption{iCISCF and iCISCF(2) energies (+994.0 $E_{\mathrm{h}}$) of hexacene with CAS(26, 26)/cc-pVDZ}
	\begin{threeparttable}
		\centering
		\makebox[\linewidth]{
			\begin{tabular}{c|cccccccccc}\toprule
				&\multicolumn{4}{c}{$^{1}A_{g}$}  & \multicolumn{4}{c}{$^{3}B_{3u}$} & \multicolumn{2}{c}{Gap (kcal/mol)} \\
				C$_{\mathrm{min}}$
				&\multicolumn{1}{c}{$N_{\mathrm{CSF}}$ \tnote{a}}
				&\multicolumn{1}{c}{$N_{\mathrm{SD}}$ \tnote{b}}
				&\multicolumn{1}{c}{$E_{\mathrm{iCISCF}}$}
				&\multicolumn{1}{c}{$E_{\mathrm{iCISCF(2)}}$}
				&\multicolumn{1}{c}{$N_{\mathrm{CSF}}$ \tnote{a}}
				&\multicolumn{1}{c}{$N_{\mathrm{SD}}$ \tnote{b}}
				&\multicolumn{1}{c}{$E_{\mathrm{iCISCF}}$}
				&\multicolumn{1}{c}{$E_{\mathrm{iCISCF(2)}}$}
				&\multicolumn{1}{c}{$E_{\mathrm{iCISCF}}$}
				&\multicolumn{1}{c}{$E_{\mathrm{iCISCF(2)}}$}\\\toprule
$1.0\times10^{-4}$     &80618   &305351         &-0.29320     &-0.30699       &46057    &200173   &-0.25549   &-0.27317   &23.66  &21.22   \\
$7.5\times10^{-5}$     &128424  &495560         &-0.29675     &-0.30901       &197610   &343609   &-0.26043   &-0.27586   &22.79  &20.80   \\
$5.0\times10^{-5}$     &227406  &900944         &-0.30084     &-0.31119       &346020   &612068   &-0.26517   &-0.27837   &22.38  &20.59   \\
$2.5\times10^{-5}$     &577599  &2405533        &-0.30641     &-0.31407       &875294   &1593875  &-0.27174   &-0.28185   &21.76  &20.22   \\
Estimated \tnote{c}    &-       &1000000        &-0.30108     &-0.31126       &-        &1000000  &-0.26847   &-0.28012   &20.46  &19.54   \\
Extrapolated \tnote{d}& $3.7\times10^{12}$ &$2.7\times10^{13}$ &-                &-0.32300       &$8.4\times10^{12}$ &$2.3\times10^{13}$&-                &-0.29345       &-     &18.53  \\
				\\
				ASCISCF(2) \tnote{e} &  -           &1000000  &-0.29562       &-0.30753       &-            &1000000  &-0.26570       &-0.27771       &18.78 &18.72  \\
				Extrapolated &  -           &-              &-                  &-              &-          &-              &-                  &-      &-     &19.4
				\\\midrule
			\end{tabular}
		}
		\begin{tablenotes}
			\item[a] Number of selected configuration state functions.
			\item[b] Number of Slater determinants corresponding to $N_{\mathrm{CSF}}$.
			\item[c] Estimated by linear fit of $E_{\mathrm{iCISCF}}$ or $E_{\mathrm{iCISCF(2)}}$ as function of $log_{10}N_{\mathrm{SD}}$.
			\item[d] $N_{\mathrm{CSF}}$/$N_{\mathrm{SD}}$: total number of CSFs/SDs in the symmetry adapted CAS.
			\item[e] Ref. \cite{ASCISCF2}.
		\end{tablenotes}
	\end{threeparttable}\label{Hexacene}
\end{table}

\begin{figure}[!htp]
	\centering
	\begin{tabular}{c}
		\includegraphics[width=0.6\textwidth]{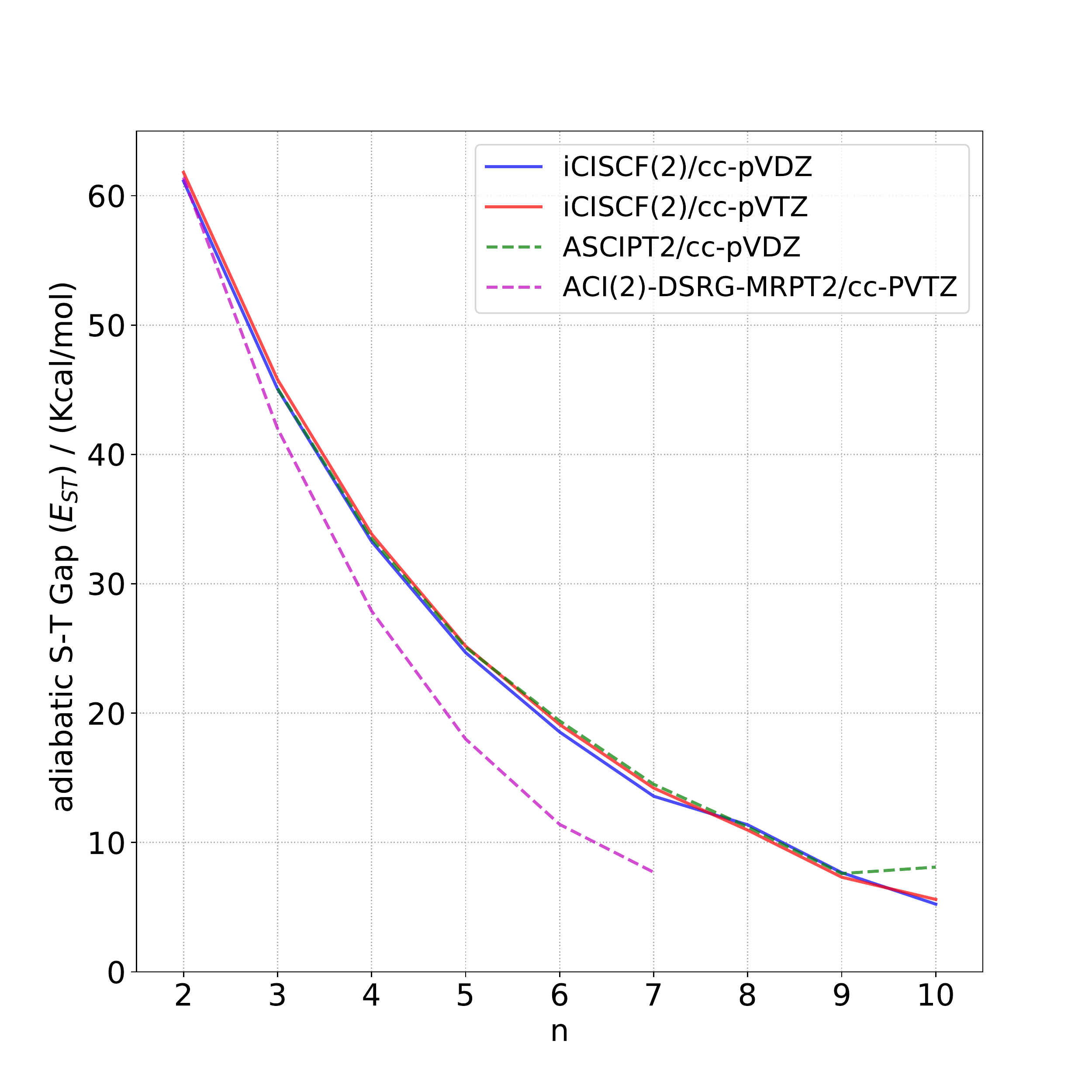}\\
		\\
	\end{tabular}
	\caption{Extrapolated iCISCF(2) adiabatic S-T gaps of polyacenes with cc-pVDZ and cc-pVTZ, to compare with those by ASCISCF(2)/cc-pVDZ\cite{ASCISCF2} and ACI(2)-DSRG-MRPT2/cc-pVTZ\cite{ACI2018}.}
	\label{Gap}
\end{figure}

\begin{figure}[!htp]
	\centering
	\begin{tabular}{c}
		\includegraphics[width=0.6\textwidth]{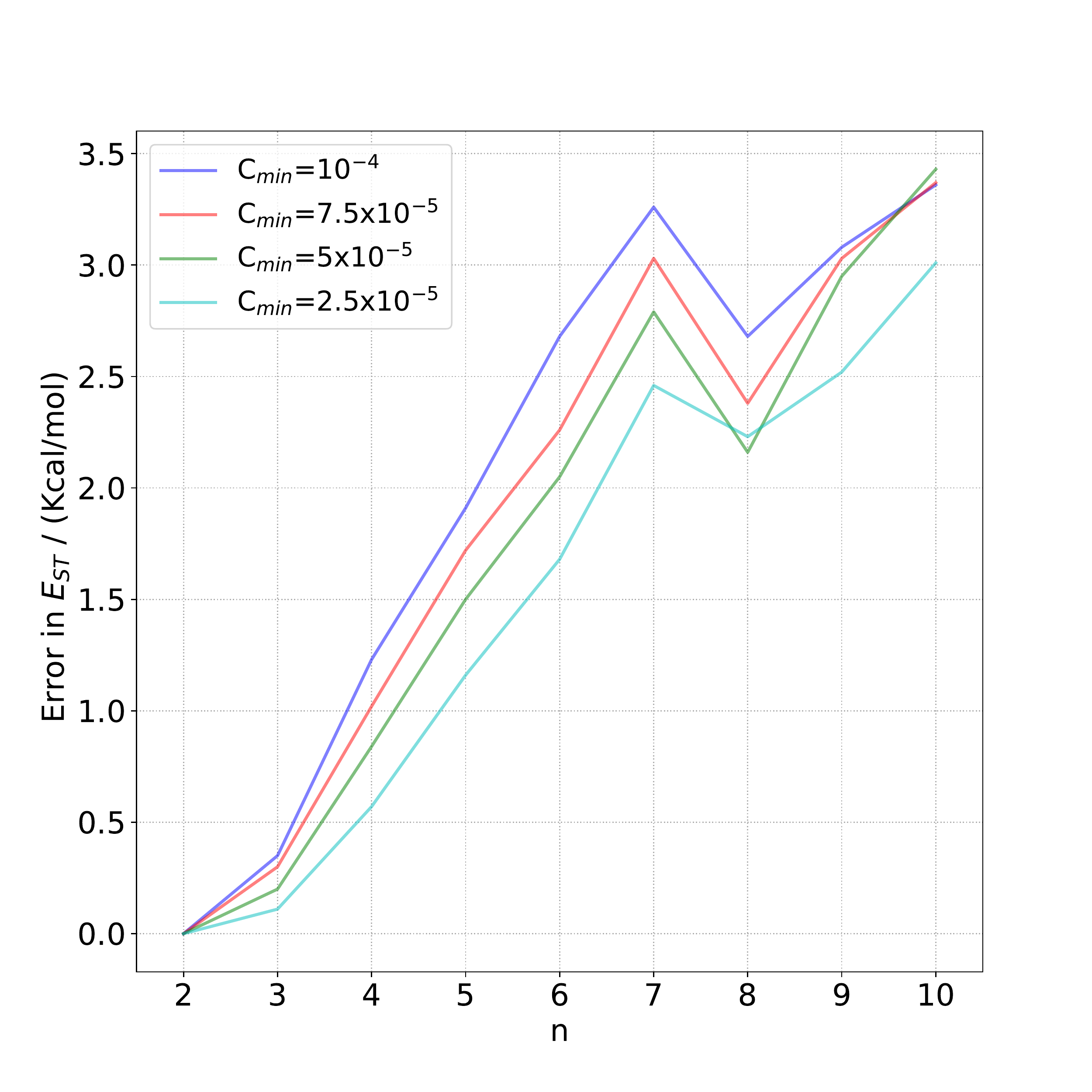}\\
		\\
	\end{tabular}
	\caption{Differences ($\Delta E_{\mathrm{ST}}$) between the calculated and extrapolated adiabatic S-T gaps of polyacenes
by iCISCF(2)/cc-pVDZ.}
	\label{Gap-DZ}
\end{figure}

The absolute energies of n-acenes computed at iCISCF and iCISCF(2) level have been provided in the Supporting information (SI). The extrapolated iCISCF(2) S-T splittings of n-acenes are plotted in Fig. \ref{Gap}. As can be seen, the results with the cc-pVDZ and cc-pVTZ basis sets are very close,
reflecting marginal basis set incompleteness errors.
Except for noacene, the iCISCF(2)/cc-pVDZ results are very similar to those by ASCISCF(2)/cc-pVDZ\cite{ASCISCF2}. The obvious deviations of iCISCF(2) (and ASCISCF(2)) from
ACI(2)-DSRG-MRPT2/cc-pVTZ (adaptive CI with second-order perturbative multireference-driven similarity renormalization group)\cite{ACI2018} should then be ascribed to dynamic correlation from the $Q$ space that is accounted for by DSRG-MRPT2. For more details,
the differences $\Delta E_{\mathrm{ST}}$ between the actually calculated and the extrapolated S-T gaps are further plotted in Fig. \ref{Gap-DZ}, which shows that
$\Delta E_{\mathrm{ST}}$ becomes larger as the system size increases but can be reduced by decreasing $C_{\mathrm{min}}$. However, even for
the smallest $C_{\mathrm{min}}$ (i.e., $2.5\times10^{-5}$) used here,
$\Delta E_{\mathrm{ST}}$ for decacene is still as large as 3.0 kcal/mol, which is to be compared with the extrapolated value of S-T gap, 5.2 kcal/mol. Therefore,
predicting accurately the S-T gaps of decacene and beyond remains a challenge if the extrapolation step is not taken.

\subsection{Fe-porphyrin complex}
\begin{figure}[!htp]
	\centering
	\begin{tabular}{c}
		\includegraphics[width=1.0\textwidth]{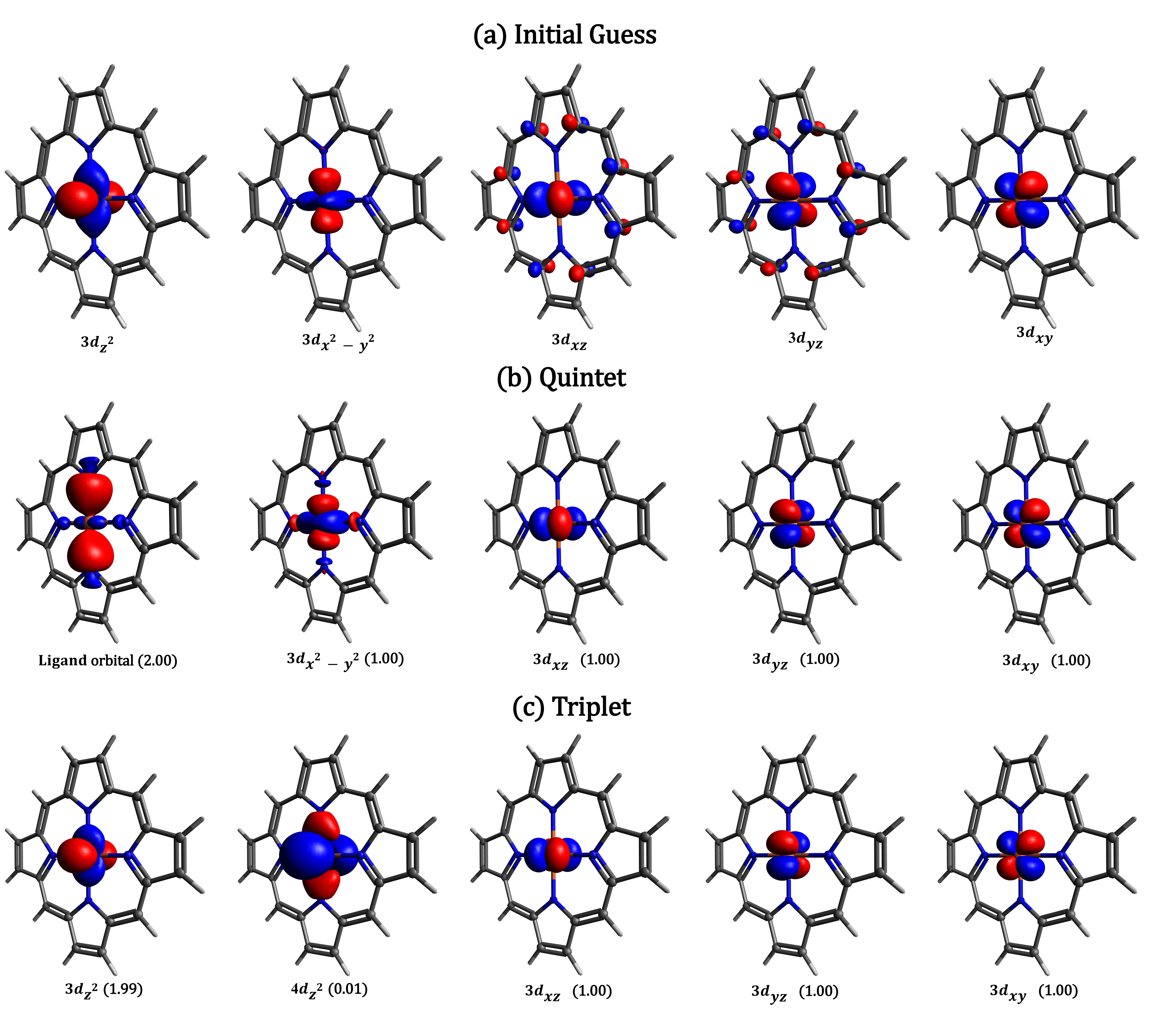}\\
		\\
	\end{tabular}
	\caption{The CAS(32, 29) active orbitals of (a) initial guess, (b) converged $^{5}A_{g}$, and (c) converged $^{3}B_{1g}$. The 24 valence $\pi$ orbitals of porphyrin are not shown.}
	\label{CAS3229}
\end{figure}

Fe(II) porphyrin (FeP) complexes
are another set of classic test systems for strongly correlated methods\cite{FeP-DMRG,HBCISCF2017,FeP-Alavi1,FeP-Alavi2,ASCISCF}, due to both large active spaces and small energy gaps between different spin states.
The free-base porphyrin already has 24 valence $\pi$ orbitals. Therefore, considering the eight electrons and five $3d$ orbitals of Fe,
a minimal active space of FeP would be as large as CAS(32, 29). Even so, the correct spin ordering still cannot be reproduced\cite{HBCISCF2017,ASCISCF}.
That $^3B_{1g}$ is lower than $^{5}A_{g}$ was only found by going to CAS(44, 44)\cite{HBCISCF2017}, which consists of additional Fe $4d4p_x4p_y$ and $4\times\mathrm{N}$ $2p_x2p_y$.
Realizing that CAS(44, 44) is not really rational,
Levine et al.\cite{ASCISCF} proposed to use CAS(40, 42), which includes Fe $4s4p4d$ and four $\sigma$ lone pair orbitals of N coordinated to Fe on top of CAS(32, 29).
To compare directly with these results, the iCISCF/cc-pVDZ calculations were performed with the $^3B_{1g}$ geometry\cite{FeP-Geom} for both $^3B_{1g}$ and $^{5}A_{g}$.
The distributions of the orbitals of FeP in the irreps of $D_{2h}$ are given in Table \ref{IrrepFeP}. 
Upon convergence after 140 macroiterations of iCISCF-CAS(32, 29) with $C_{\mathrm{min}}=1.0\times10^{-4}$, close inspections reveal that
24 of the 29 active orbitals are indeed the valence $\pi$ orbitals of porphyrin, but the remaining 5 are not all the expected Fe 3$d$.
It can be seen clearly from Fig. \ref{CAS3229} that in the case of $^{5}A_{g}$, the nearly doubly occupied 3$d_{z^2}$ orbital is pushed out of the active space, whereas
a doubly occupied ligand orbital is rotated into the active space to correlate the singly occupied 3$d_{x^2-y^2}$ orbital.
For the $^{3}B_{1g}$ state, the converged active space does not contain the 3$d_{x^2-y^2}$ orbital, which is pushed to the virtual space.
Instead, the 4$d_{z^2}$ orbital enters the active space to incorporate the 3$d_{z^2}$ orbital.
The five 3$d$ orbitals can be kept inside the active space only by freezing the 3$d_{z^2}$ (3$d_{x^2-y^2}$) orbital in the
calculations of $^{5}A_{g}$ ($^{3}B_{1g}$).

\begin{table}[!htp]
	\small
	\centering
	\caption{ Distributions of orbitals of FeP in irreducible representations of $D_{2h}$}
	\begin{threeparttable}
		\centering
		\begin{tabular}{cc|cccccccc}\toprule			
			 \multicolumn{1}{c}{active space}			
			&\multicolumn{1}{c}{$$}
			&\multicolumn{1}{c}{$A_{g}$}
			&\multicolumn{1}{c}{$B_{3g}$}
			&\multicolumn{1}{c}{$B_{2g}$}
			&\multicolumn{1}{c}{$B_{1g}$}
			&\multicolumn{1}{c}{$A_{u}$}
			&\multicolumn{1}{c}{$B_{3u}$}
			&\multicolumn{1}{c}{$B_{2u}$}
			&\multicolumn{1}{c}{$B_{1u}$}\\\toprule
			\multirow{2}{*}{(32, 29)}
            & core	&23	&0	&0	&14	&0	&19	&19	&2  \\
            & active	&2	&7	&7	&1	&5	&0	&0	&7  \\   			
			\multirow{2}{*}{(40, 42)}
			& core	&21	&0	&0	&14	&0	&18	&18	&2  \\
			& active	&7	&8	&8	&2	&5	&2	&2	&8  \\
			\multirow{2}{*}{(44, 44)}
			& core	&21	&0	&0	&14	&0	&17	&17	&2  \\
			& active	&6	&8	&8	&4	&5	&3	&3	&7  \\\toprule	
		\end{tabular}
	\end{threeparttable}\label{IrrepFeP}
\end{table}

The iCISCF-CAS(32, 29) energies are documented in Table \ref{FeP3229}.
As the previous studies\cite{HBCISCF2017,ASCISCF}, the $^{5}A_{g}$ state is predicted to be the ground state.
The extrapolated iCISCF(2) results are in good agreement with those from HCISCF(2)\cite{HBCISCF2017}
for both the total and relative energies. In contrast, the extrapolated energy gap by ASCISCF(2)\cite{ASCISCF} is just half that
by iCISCF(2) or HCISCF(2). This is mainly because the ASCISCF(2) energy for $^{3}B_{1g}$ is too low as compared with the iCISCF(2) one.
Freezing the  3$d_{x^2-y^2}$  (3$d_{z^2}$)
orbital inside the active space increases the energy of $^{3}B_{1g}$ ($^{5}A_{g}$) by 3.6 (0.3) kcal/mol, thereby increasing the gap by 3.3 kcal/mol.

\begin{table}[!htp]
	\footnotesize
	\centering
	\caption{iCISCF-CAS(32, 29) energies (+2245.0 $E_{\mathrm{h}}$) of FeP $^{\mathrm{a}}$}
	\begin{threeparttable}
		\centering
		\makebox[\linewidth]{
			\begin{tabular}{c|cccccccccc}\toprule
				&\multicolumn{4}{c}{$^{3}B_{1g}$}  & \multicolumn{4}{c}{$^{5}A_{g}$} &
				\multicolumn{2}{c}{Gap (kcal/mol)} \\
				$C_{\mathrm{min}}$
				&\multicolumn{1}{c}{$N_{\mathrm{CSF}}$}
				&\multicolumn{1}{c}{$N_{\mathrm{SD}}$}
				&\multicolumn{1}{c}{$E_{\mathrm{iCISCF}}$}
				&\multicolumn{1}{c}{$E_{\mathrm{iCISCF(2)}}$}
				&\multicolumn{1}{c}{$N_{\mathrm{CSF}}$}
				&\multicolumn{1}{c}{$N_{\mathrm{SD}}$}
				&\multicolumn{1}{c}{$E_{\mathrm{iCISCF}}$}
				&\multicolumn{1}{c}{$E_{\mathrm{iCISCF(2)}}$}
				&\multicolumn{1}{c}{$E_{\mathrm{iCISCF}}$}
				&\multicolumn{1}{c}{$E_{\mathrm{iCISCF(2)}}$}\\\toprule
				&&&&&frozen \tnote{b}&&&&\\
				$1.0\times10^{-4}$     &  103238 & 186387 & 0.02457     &       0.00983   &108153 &161556   &-0.00508   &-0.02113       &-18.61  &-19.43   \\
				$7.5\times10^{-5}$     &  154932 & 284279 & 0.02144     &       0.00850   &165285 &249441   &-0.00860   &-0.02283       &-18.85  &-19.66   \\
				$5.0\times10^{-5}$     &  271010 & 507312 & 0.01768     &       0.00691   &296625 &454698   &-0.01231   &-0.02413       &-18.82  &-19.48   \\
				$2.5\times10^{-5}$     &  691268 & 1331168& 0.01258     & 0.00474   &774101 &1211597    &-0.01777       &-0.02639       &-19.04  &-19.53   \\
				Extrapolated &  $1.7\times10^{14}$& $5.0\times10^{14}$& -         &-0.0010    & $1.7\times10^{14}$& $3.4\times10^{14}$&   -       &-0.0323    &-       &-19.64   \\
				\\
				&&&&\multicolumn{3}{c}{fully optimized}&&&\\
				$1.0\times10^{-4}$     &  106323 & 193856  & 0.02069    &0.00467          &109165        &162983        &-0.00533       &-0.02149        &-16.33  &-16.41  \\
				$7.5\times10^{-5}$     &  165932 & 307657  & 0.01723    &0.00323          &165971        &250522        &-0.00858       &-0.02284        &-16.19  &-16.36  \\
				$5.0\times10^{-5}$     &  296081 & 558728  & 0.01314    &0.00152          &302190        &463045        &-0.01273       &-0.02455        &-16.23  &-16.36  \\
				$2.5\times10^{-5}$     &  758072 & 1469095 & 0.00763    &-0.00077   &785566      &1228691       &-0.01818       &-0.02680        &-16.19  &-16.33  \\
				Extrapolated  &  $1.7\times10^{14}$& $5.0\times10^{14}$ & -        & -0.0068   &  $1.7\times10^{14}$& $3.4\times10^{14}$ &    -     &-0.0328      &-       &-16.36  \\
				\\
				HCISCF(2) \tnote{c}   &  -      &533623  &0.0224     &-            & -          &379536    &0.0020           &-            &-12.80  &-        \\
				Extrapolated          &  -           &-       &  -            &-0.0049          & -          &-       &-          &-0.0314        &-       &-16.63   \\
				\\
				ASCISCF(2) \tnote{d}  &  -      &500000     &-0.0022       &-              & -          &500000    &-0.0191          &-            &-10.60  &-        \\
				Extrapolated          &  -           &-        &  -           &-0.0186    & -        &-       &-                &-0.0324          &-       &-8.63    \\\midrule
			\end{tabular}
		}
		\begin{tablenotes}
			\item[a] See Table \ref{Hexacene} for additional explanations.
			\item[b] The 3$d_{z^2}$ and 3$d_{x^2-y^2}$ orbitals are kept frozen in the active space when calculating $^{5}A_{g}$ and $^{3}B_{1g}$, respectively.
			\item[c] Ref. \cite{HBCISCF2017}
			\item[d] Ref. \cite{ASCISCF}
		\end{tablenotes}
	\end{threeparttable}\label{FeP3229}
\end{table}

To compare with the HCISCF-CAS(44, 44) results\cite{HBCISCF2017}, the iCISCF-CAS(44, 44) calculations were also performed. Except for the 24 $\pi$ orbitals of porphyrin,
the remaining active orbitals are plotted in Fig. S1 in the SI. It turns out that the Fe 4$d_{x^2-y^2}$ orbital is pushed out of
the active space (in line with the previous observation\cite{ASCISCF}), while the 27$^{th} A_g$ orbital, consisting mainly of Fe $4s$ and N $3p$, is pulled into the active space.
The former can be viewed as the  ``anti-bonding'' orbital of 3$d_{x^2-y^2}$ whereas the latter can be viewed as the ``anti-bonding''
orbital of the 22$^{th} A_g$, a lone pair orbital of four N atoms coordinate to Fe. It is better to put both 4$d_{x^2-y^2}$ and $4s$ into the active space as in the case of CAS(40, 42).
If only one of them is to be included, the Fe $4s$ instead of 4$d_{x^2-y^2}$ orbital is favored for lower energies.
Again, we performed calculations by freezing the 4$d_{x^2-y^2}$ orbital in the active space, see Fig. S2 for the converged 4$d_{x^2-y^2}$ and 4$d_{z^2}$ orbitals.
The latter is slightly mixed with the 4$s$ orbitals after optimization.

\begin{table}[!htp]
	\footnotesize
	\centering
	\caption{iCISCF-CAS(44, 44) energies (+2245.0 $E_{\mathrm{h}}$) of FeP $^{\mathrm{a}}$}
	\begin{threeparttable}
		\centering
		\makebox[\linewidth]{
			\begin{tabular}{c|cccccccccc}\toprule
				&\multicolumn{4}{c}{$^{3}B_{1g}$}  & \multicolumn{4}{c}{$^{5}A_{g}$} &
				\multicolumn{2}{c}{Gap (kcal/mol)} \\
				C$_{\mathrm{min}}$
				&\multicolumn{1}{c}{$N_{\mathrm{CSF}}$}
				&\multicolumn{1}{c}{$N_{\mathrm{SD}}$}
				&\multicolumn{1}{c}{$E_{\mathrm{iCISCF}}$}
				&\multicolumn{1}{c}{$E_{\mathrm{iCISCF(2)}}$}
				&\multicolumn{1}{c}{$N_{\mathrm{CSF}}$}
				&\multicolumn{1}{c}{$N_{\mathrm{SD}}$ }
				&\multicolumn{1}{c}{$E_{\mathrm{iCISCF}}$}
				&\multicolumn{1}{c}{$E_{\mathrm{iCISCF(2)}}$}
				&\multicolumn{1}{c}{$E_{\mathrm{iCISCF}}$}
				&\multicolumn{1}{c}{$E_{\mathrm{iCISCF(2)}}$}\\\toprule
				&&&&&frozen \tnote{b}&&&&\\
				$1.0\times10^{-4}$     &197817  &366295   &-0.133538    &-0.169007      &179562   &264701        &-0.135383     &-0.167162      &-1.16 &1.16 \\
				$7.5\times10^{-5}$     &309455  &582411   &-0.139264    &-0.171517      &280883   &419065        &-0.140490 &-0.169338  &-0.77 &1.37 \\
				$5.0\times10^{-5}$     &574704  &1103405        &-0.146524      &-0.174667      &527665   &800161        &-0.147107     &-0.172201      &-0.37 &1.55 \\
				$2.5\times10^{-5}$     &1629683 &3222519        &-0.156921      &-0.179173      &1510972        &2345179 &-0.156565     &-0.176247      &0.22  &1.84 \\
				Extrapolated&$1.2\times10^{23}$& $5.1\times10^{23}$&-          &-0.1963
				&$1.4\times10^{23}$& $3.9\times10^{23}$&-         &-0.1911      &-     &3.28 \\
				\\
				&&&&\multicolumn{3}{c}{fully optimized}&&&\\
				$1.0\times10^{-4}$     &199609  &372869 &-0.15415         &-0.19010     &180638   &267708         &-0.14558     &-0.17787       &5.38 &7.67 \\
				$7.5\times10^{-5}$     &317682  &603672 &-0.16021         &-0.19264     &281562   &422181         &-0.15082     &-0.18003       &5.90 &7.91 \\
				$5.0\times10^{-5}$     &587999  &1138209        &-0.16752       &-0.19583       &543338   &828687         &-0.15784     &-0.18304       &6.07 &8.02 \\
				$2.5\times10^{-5}$     &1648308 &3278017        &-0.17797       &-0.20037       &1541226        &2402169        &-0.16738       &-0.18716       &6.64 &8.29 \\
				Extrapolated&$1.2\times10^{23}$& $5.1\times10^{23}$&-            &-0.2174        &$1.4\times10^{23}$& $3.9\times10^{23}$&-  &-0.2019          &-    &9.73 \\
				\\
				HCISCF(2) \tnote{c}      &-              &2133424  &-0.1567          &-            &-            &1450271  &-0.1457  &-            &6.90  &-    \\
				Extrapolated          &-              &-              &-                &-0.1996          &-          &-              &-              &-0.1965    &-     &1.93 \\
				\\
				
				\\
				ASCISCF(2) \tnote{d}  & -&500000 & -0.2044 & - &- & 500000 & -0.2315 & - &-17.01 &-\\
				Extrapolated          &-              &-              &-                &-0.25617          &-          &-              &-              &-0.28688    &-     &-19.27 
				\\\midrule
			\end{tabular}
		}
		\begin{tablenotes}
			\item[a] See Table \ref{Hexacene} for additional explanations.
			\item[b] The 4$d_{x^2-y^2}$ orbitals are kept frozen in the active space.
			\item[c] Ref. \cite{HBCISCF2017}
			\item[d] Ref. \cite{ASCISCF}
		\end{tablenotes}
	\end{threeparttable}\label{FeP4444}
\end{table}

The iCISCF-CAS(44, 44) energies are documented in Table \ref{FeP4444}. In all calculations with different $C_{\mathrm{min}}$, the $^3B_{1g}$ state is predicted to be the ground state.
However, the presently extrapolated iCISCF(2) energy gap (9.7 kcal/mol) between $^{5}A_{g}$ and $^3B_{1g}$ is significantly larger than that (1.9 kcal/mol) by HCISCF(2)\cite{HBCISCF2017}.
The latter is even smaller than the gap (3.3 kcal/mol) obtained by iCISCF(2) with 4$d_{x^2-y^2}$ frozen. Close inspections reveal that,
although the HCISCF gap (6.9 kcal/mol) is very much the same as the iCISCF one (6.6 kcal/mol) with $C_{\mathrm{min}}=2.5\times 10^{-5}$, the HCISCF energies for
$^{5}A_{g}$ and $^3B_{1g}$ are actually close to the iCISCF ones with $C_{\mathrm{min}}=1.0\times 10^{-4}$, given dramatic differences in the numbers of SDs. It can therefore be concluded
that the HCISCF calculations\cite{HBCISCF2017} are far from convergence.
Note also that ASCISCF-CAS(44, 44)\cite{ASCISCF} predicts much lower variational energies but still incorrect spin ordering, which was ascribed to
spatial symmetry breaking. To investigate such symmetry-breaking effects, we repeated the calculations using C$_{2v}$ instead of D$_{2h}$. Indeed,
localized symmetry-breaking active orbitals are observed (cf. Fig. S3), leading to lower energies than those with D$_{2h}$.

Some representative active orbitals from iCISCF-CAS(40, 42) calculations are plotted in Fig. S4. As can be seen from the energetics in
Table \ref{FeP4042}, both iCISCF and iCISCF(2) predict the correct ground state with different $C_{\mathrm{min}}$. And in this case,
there exist perfect agreements between iCISCF and ASCISCF\cite{ASCISCF} for both the extrapolated total and relative energies. Yet,
rather unexpectedly, the iCISCF-CAS(40, 42) energies are even lower than the
iCISCF-CAS(44, 44) ones (cf. Table \ref{FeP4444}), reflecting the importance of rational construction of active spaces (for which the $\mathbbm{i}$CAS method is a good choice).

\begin{table}[!htp]
	\footnotesize
	\centering
	\caption{iCISCF-CAS(40, 42) energies (+2245.0 $E_{\mathrm{h}}$) of FeP $^{\mathrm{a}}$}
	\begin{threeparttable}
		\centering
		\makebox[\linewidth]{
			\begin{tabular}{c|cccccccccc}\toprule
				&\multicolumn{4}{c}{$^{3}B_{1g}$}  & \multicolumn{4}{c}{$^{5}A_{g}$} &
				\multicolumn{2}{c}{Gap (kcal/mol)} \\
				C$_{\mathrm{min}}$
				&\multicolumn{1}{c}{$N_{\mathrm{CSF}}$}
				&\multicolumn{1}{c}{$N_{\mathrm{SD}}$$^a$}
				&\multicolumn{1}{c}{$E_{\mathrm{iCISCF}}$}
				&\multicolumn{1}{c}{$E_{\mathrm{iCISCF(2)}}$}
				&\multicolumn{1}{c}{$N_{\mathrm{CSF}}$}
				&\multicolumn{1}{c}{$N_{\mathrm{SD}}$$^a$}
				&\multicolumn{1}{c}{$E_{\mathrm{iCISCF}}$}
				&\multicolumn{1}{c}{$E_{\mathrm{iCISCF(2)}}$}
				&\multicolumn{1}{c}{$E_{\mathrm{iCISCF}}$}
				&\multicolumn{1}{c}{$E_{\mathrm{iCISCF(2)}}$}\\\toprule
				$1.0\times10^{-4}$     &199810  &370332  &-0.15477      &-0.19185 &176251         &259571        &-0.15428      &-0.18806 &0.31 &2.38 \\
				$7.5\times10^{-5}$     &314006  &591402  &-0.16086      &-0.19468 &279206         &416350        &-0.16002      &-0.19072 &0.52 &2.48 \\
				$5.0\times10^{-5}$     &595172  &1143717 &-0.16882      &-0.19838 &535154         &811146        &-0.16738      &-0.19412 &0.91 &2.67 \\
				$2.5\times10^{-5}$     &1686557 &3331667 &-0.18018      &-0.20369 &1566899       &2429062    &-0.17800  &-0.19907 &1.37 &2.90 \\
				Estimated   & -      & 500000 &-0.1587   & -       & -        & 500000    & -0.1617  & -       &-1.87 &- \\
				Extrapolated           &$7.3\times10^{21}$&$3.0\times10^{22}$       & -        &-0.2242 &$8.5\times10^{21}$& $2.2\times10^{22}$ &-         &-0.2173  & -    &4.34 \\
				\\
				ASCISCF(2) \tnote{b}  &-       &500000      &-0.1671       &-            &-              &500000    &-0.1699       &-            &-1.76 &-\\
				Extrapolated           &-           &-       &-         &-0.2208        &-            &-             &-         &-0.2137  & -    &4.46\\\midrule
			\end{tabular}
		}
		\begin{tablenotes}
			\item[a] See Table \ref{Hexacene} for additional explanations.
			\item[b] Ref. \cite{ASCISCF}
		\end{tablenotes}
	\end{threeparttable}\label{FeP4042}
\end{table}

\subsection{A rational design of active space for Fe-porphyrin}
Although the FeP complexes have been studied by many groups utilizing different active spaces, there are not many works explaining how their active spaces for FeP are designed. It has been proved that the $\mathbbm{i}$CAS method developed by some of us is a useful tool to design and generate initial guess MOs for MCSCF calculations.\cite{iCAS} In this section, we would like to design a few active spaces from small to large utilizing the $\mathbbm{i}$CAS algorithm. All initial guess MOs in this section are generated based on the ROHF wave function of $^{5}A_{g}$ state. 

\begin{figure}[!htp]
	\centering
	\begin{tabular}{c}
		\includegraphics[width=1.0\textwidth]{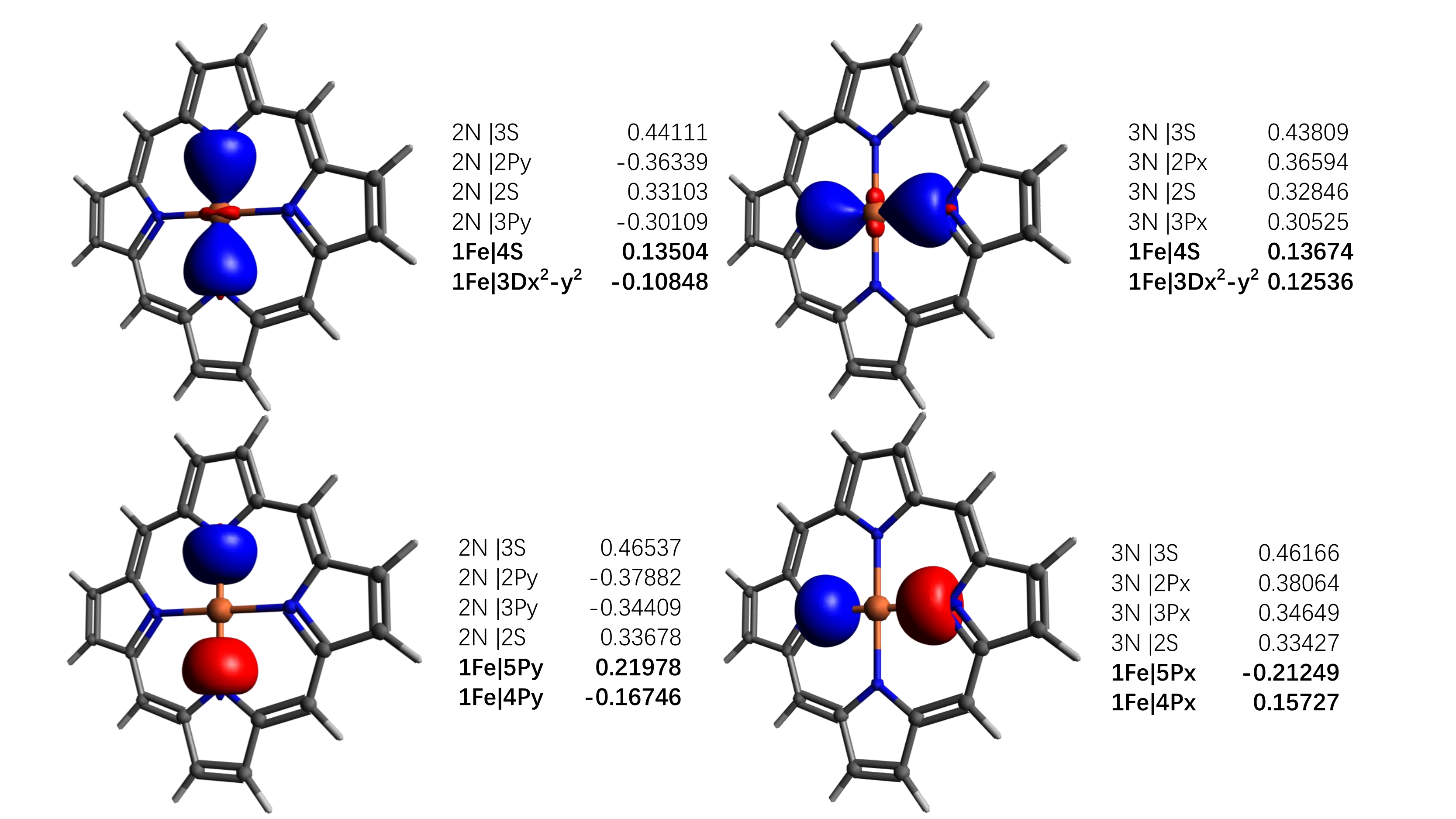}\\
		\\
	\end{tabular}
	\caption{The four ligand orbitals close to the Fe centre. Their significant MO coefficients are listed as well.}
	\label{ligand}
\end{figure}

One of the smallest active space to describe various spin states of FeP is CAS(6, 5), which contains only the 3$d$ orbitals and electrons of Fe$^{2+}$. It has been proved that this small active space is not able to describe the strong interaction between Fe$^{2+}$ and porphyrin anion.\cite{FeP-PIERLOOT1} More orbitals and electrons have to be included in the active space. By comparing the results of CAS(32, 29) and CAS(40, 42) in Sec. 3.3, it is evident that the $\pi$ orbitals of porphyrin do not influence the spin splitting of Fe$^{2+}$ as significantly as the ligand orbitals and 4$s$4$p$4$d$ orbitals of Fe do. Usually, the ligand orbitals interact with the metal centre strongly. Thus, in the first step, the five 3$d$ orbitals of Fe, the 2$s$, 2$p_x$, 2$p_y$ orbitals of the four N atoms are used to generate pre-LMOs and the subsequent LMOs guesses for iCISCF calculations. Besides the five 3$d$ orbitals, 12 LMOs, eight doubly occupied MOs (DOMOs) and four virtual MOs, are generated by the $\mathbbm{i}$CAS as active MOs. In fact, not all the 12 LMOs are spatially close to the Fe$^{2+}$. By checking their coefficients, we found that only four doubly occupied LMOs are mixed with the atomic orbitals of Fe. The four ligand LMOs are exhibited in Fig. \ref{ligand}, together with their significant MOs coefficients. If only the four LMOs and the Fe 3$d$ orbitals are considered as active MOs, the active space will be CAS(14,9). However, such an active space only contains DOMOs and SOMOs, and does not have any virtual orbitals to relax the electrons in the active space. By analyzing the coefficients of the four LMOs, another 3 atomic orbitals are added, the 4$s$, 4$p_x$ and 4$p_y$ orbtial of Fe$^{2+}$. Thus, a CAS(14,12) active space is generated by the $\mathbbm{i}$CAS method. Moreover, to account for the doubles-shell effects, the 4$d$ orbitals of Fe atom could be considered, resulting in iCISCF-CAS(14,17) calculations.

\begin{figure}[!htp]
	\centering
	\begin{tabular}{c}
		\includegraphics[width=0.8\textwidth]{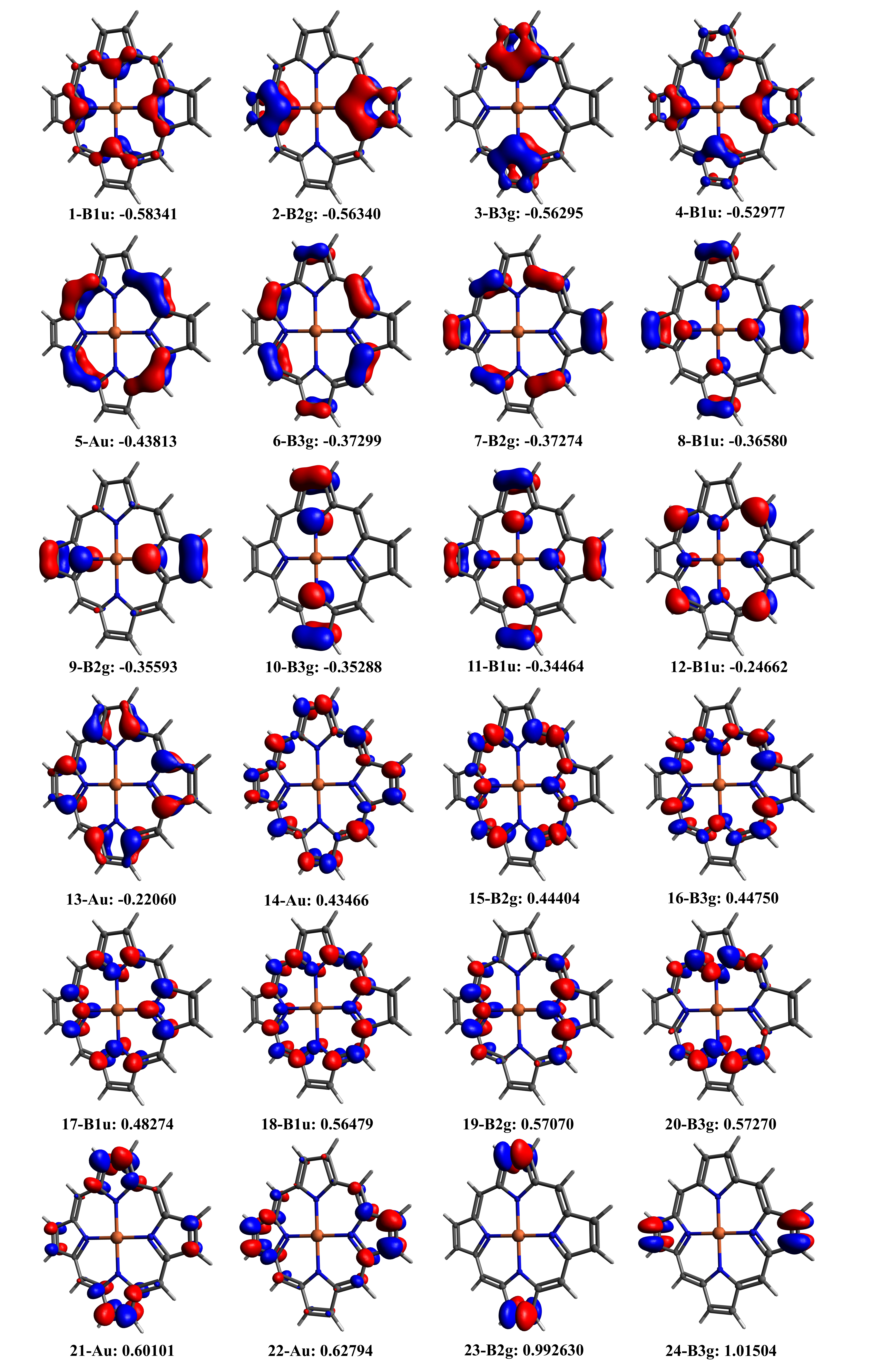}\\
		\\
	\end{tabular}
	\caption{The 24 $\pi$ regional LMOs of porphyrin constructed by the $\mathbbm{i}$CAS method, ordered by their orbital energies (in $E_{\mathrm{h}}$).}
	\label{PIMO}
\end{figure}

Although the $\pi$ orbitals of porphyrin are not spatially adjacent to Fe$^{2+}$, some studies showed that they do differentially stabilize the triplet states over quintet states of FeP.\cite{FeP-Alavi3} For the low-lying excited states of FeP, the 24 $\pi$ orbitals are distributed to four different irreps of $D_{2h}$ symmetry, six $B_{3g}$ MOs (three DOMOs), six $B_{2g}$ MOs (three DOMOs), five $A_{u}$ MOs (two DOMOs), and seven $B_{1u}$ MOs (five DOMOs). To consider the influence of the $\pi$ orbitals, it may not be necessary to include all 24 $\pi$ MOs in the active space. In principle, the $\pi$ orbitals with close energies as the HOMO and LUMO, should interact with the Fe$^{2+}$ more strongly. Thus, the 24 $\pi$ LMOs, constructed from the pre-LMOs, are recanonicalized based on their Fock matrix (the Fock matrix of DOMOs or virtual MOs are diagonalized separately), which is called regional LMOs. After the canonicalization, the eigenvalues of the Fock matrix within each subspace can be considered as approximate orbital energies. The 24 $\pi$ regional LMOs with their irreps and energies are given in Fig. \ref{PIMO}. Thus, for the first step, ten orbitals (from the 9$^{th}$ to 18$^{th}$ MOs in Fig. \ref{PIMO}) are added to the CAS(14,17) active space, including two $B_{3g}$ MOs (one DOMO), two $B_{2g}$ MOs (one DOMO), two $A_{u}$ MOs (one DOMO), and four $B_{1u}$ MOs (two DOMOs). Moreover, we found that the orbital dominated by the 4$p_z$ orbital of Fe with $B_{1u}$ symmetry has lower orbital energy than the frontier virtual $\pi$ orbital in Fig. \ref{PIMO} (the 14$^{th}$ MO), which could be included in the active space as well. Thus, two sets of guess MOs with and without the 4$p_z$ orbital of Fe centre, CAS(14,28) and CAS(14,27), are generated as well. To consider the influence of more $\pi$ orbitals, another seven orbitals (from the 5$^{th}$ to 8$^{th}$ and from 19$^{th}$ to 21$^{th}$ in Fig. \ref{PIMO}) could be taken into account. Thus, on top of CAS(24,28), eight more electrons and seven more active MOs are added to the active space, resulting in iCISCF-CAS(32,35) calculations. Finally, by adding the remaining valence $\pi$ orbital of porphyrin to the active space, we arrive at the active space proposed by Levine and coworkers, CAS(40,42). The distribution of the core and active orbitals of calculations with the above mentioned seven different active spaces are given in Table \ref{IrrepFeP2}.

Using all active spaces in Table \ref{IrrepFeP2}, the $^3A_{2g}$ ($^3B_{1g}$ in $D_{2h}$), $^3E_{g}$ ($^3B_{2g}$+$^3B_{3g}$ in $D_{2h}$) and $^5A_{1g}$ ($^5A_{g}$ in $D_{2h}$) state of FeP are computed. Note that for the Cartesian coordinates offered by Groenhof and coworkers, the $D_{4h}$ symmetry is not applied during the geometry optimization\cite{FeP-Geom}. Thus, the degenerate $^3E_{g}$ state is split to two states in $D_{2h}$ symmetry, $^3B_{2g}$ and $^3B_{3g}$. The three triplet and one quintet states computed using various active space are given in Table \ref{FePCAS}. The results show that with the two smallest active space, CAS(6,5) and CAS(14,12), the $^5A_{g}$ state is predicted to be the ground state. Thus, to include only the ligand orbitals in the active space, the triplet states are not stabilized over the quintet state very much. However, by taking into account the double-shell effects (iCISCF-CAS(14,17) results in Table \ref{IrrepFeP2}), the $^5A_{g}-^3B_{1g}$ splitting amounts to -0.55 kcal/mol. Compared to that with CAS(14,12), the $^5A_{g}-^3B_{1g}$ gap increases 13.90 kcal/mol. Moreover, the excitation energies of $^3B_{2g}$ and $^3B_{3g}$ state with respect to the $^3B_{1g}$ state are close to that delivered by the iCISCF-CAS(40,42) calculations as well. Thus, it is the five 4$d$ orbitals destabilizing the quintet states with respect to the $^3B_{1g}$ state more significantly.

To include some $\pi$ orbitals of porphyrin or the 4$p_z$ orbital of Fe in the active space, iCISCF results using four different active spaces are reported in Table \ref{IrrepFeP2} as well. Compared to CAS(24,27), the relative energies computed by iCISCF-CAS(24,28) are in good agreement with the most accurate ones, iCISCF-CAS(40,42). The deviations of excitation energies computed by the extrapolated iCISCF results are less than 1.0 kcal/mol. Further increase the active space to CAS(32,35), more accurate $^5A_{g}-^3B_{1g}$ gaps are predicted. However, the relative energies of $^3B_{2g}$ and $^3B_{3g}$ states with respect to the $^3B_{1g}$ state are getting worse. Nevertheless, the absolute deviations of extrapolated results are still less than 1.0 kcal/mol with respect to the iCISCF-CAS(40,42) results. 

The results in Table \ref{IrrepFeP2} show that the 4$d$ orbitals are the most important to differentially stabilize the triplet states over the $^5A_{g}$ state. Although the $\pi$ orbitals of porphyrin are essential to deliver correct spin ordering, it may not be necessary to take all the 24 $\pi$ orbitals into consideration. The active space CAS(24,28) can already predict qualitatively correct results for the four states, which could be used as starting points for the subsequent dynamic calculations.                   

\begin{table}[!htp]
	\small
	\centering
	\caption{ Distributions of orbitals of FeP in irreducible representations of $D_{2h}$ for some iCISCF calculations}
	\begin{threeparttable}
		\centering
		\begin{tabular}{cc|cccccccc}\toprule			
			\multicolumn{1}{c}{active space}			
			&\multicolumn{1}{c}{$$}
			&\multicolumn{1}{c}{$A_{g}$}
			&\multicolumn{1}{c}{$B_{3g}$}
			&\multicolumn{1}{c}{$B_{2g}$}
			&\multicolumn{1}{c}{$B_{1g}$}
			&\multicolumn{1}{c}{$A_{u}$}
			&\multicolumn{1}{c}{$B_{3u}$}
			&\multicolumn{1}{c}{$B_{2u}$}
			&\multicolumn{1}{c}{$B_{1u}$}\\\toprule
\multirow{2}{*}{(\; 6, \;5)}
& core  	&23	&3	&3	&14	&2	&19	&19	&7  \\
& active	&2	&1	&1	&1	&0	&0	&0	&0  \\
\multirow{2}{*}{(14, 12)}
& core  	&21	&3	&3	&14	&2	&18	&18	&7  \\
& active	&5	&1	&1	&1	&0	&2	&2	&0  \\
\multirow{2}{*}{(14, 17)}
& core  	&21	&3	&3	&14	&2	&18	&18	&7  \\
& active	&7	&2	&2	&2	&0	&2	&2	&0  \\
\multirow{2}{*}{(24, 27)}
& core  	&21	&2	&2	&14	&1	&18	&18	&5  \\
& active	&7	&4	&4	&2	&2	&2	&2	&4  \\   			
\multirow{2}{*}{(24, 28)}
& core  	&21	&2	&2	&14	&1	&18	&18	&5  \\
& active	&7	&4	&4	&2	&2	&2	&2	&5  \\ 
\multirow{2}{*}{(32, 35)}
& core  	&21	&1	&1	&14	&0	&18	&18	&4  \\
& active	&7	&6	&6	&2	&4	&2	&2	&6  \\
\multirow{2}{*}{(40, 42)}
& core	    &21	&0	&0	&14	&0	&18	&18	&2  \\
& active	&7	&8	&8	&2	&5	&2	&2	&8  \\\toprule	
		\end{tabular}
	\end{threeparttable}\label{IrrepFeP2}
\end{table}

\begin{table}[!htp]
	\footnotesize
	\centering
	\caption{ The energies of lowest $^3B_{1g}$ state of FeP (in $E_{\mathrm{h}}$) using seven different active spaces, the relative energies of $^3B_{2g}$, $^3B_{3g}$, and $^5A_{g}$ state w.r.t the $^3B_{1g}$ state are given in kcal/mol}
	\begin{threeparttable}
		\centering
		\makebox[\linewidth]{
			\begin{tabular}{c|cccccccc}\toprule
				&\multicolumn{4}{c}{$E_{\mathrm{iCISCF}}$}
				&\multicolumn{4}{c}{$E_{\mathrm{iCISCF(2)}}$}\\
C$_{\mathrm{min}}$
&$^3B_{1g}$ &$^3B_{2g}$ &$^3B_{3g}$ &$^5A_{g}$
&$^3B_{1g}$ &$^3B_{2g}$ &$^3B_{3g}$ &$^5A_{g}$\\\toprule
&&&&\multicolumn{2}{c}{CAS(6,5)}&&&\\
$1.0\times10^{-4}$     &-2244.72472   &4.33 	&4.19 	&-17.55   &-2244.72472   &4.33 	&4.19 	&-17.55   \\
$0.0$                 &-2244.72472   &4.33 	&4.19 	&-17.55   &-2244.72472   &4.33 	&4.19 	&-17.55   \\
&&&&&&&&\\
&&&&\multicolumn{2}{c}{CAS(14,12)}&&&\\
$1.0\times10^{-4}$     &-2244.76054    &-0.06 	&-0.24 	&-14.45    &-2244.76057    &-0.08 	&-0.26 	&-14.45    \\
$0.0$                 &-2244.76057    &-0.08 	&-0.26 	&-14.45    &-2244.76057    &-0.08 	&-0.26 	&-14.45    \\
&&&&&&&&\\
&&&&\multicolumn{2}{c}{CAS(14,17)}&&&\\
$1.0\times10^{-4}$     &-2244.86467 &1.94 	&1.68 	&-0.74  &-2244.86554 &1.80 	&1.54 	&-0.58 \\
$7.5\times10^{-5}$     &-2244.86487 &1.90 	&1.63 	&-0.67  &-2244.86556 &1.80 	&1.54 	&-0.57 \\
$5.0\times10^{-5}$     &-2244.86507 &1.86 	&1.60 	&-0.65  &-2244.86558 &1.80 	&1.53 	&-0.57 \\
$2.5\times10^{-5}$     &-2244.86532 &1.84 	&1.58 	&-0.60  &-2244.86559 &1.80 	&1.53 	&-0.57 \\
Extrapolated           &        -    &  -     &    -   &     -  &-2244.86562 &1.79 	&1.54 	&-0.55 \\
&&&&&&&&\\
&&&&\multicolumn{2}{c}{CAS(24,27)}&&&\\
$1.0\times10^{-4}$   &-2244.98960   &2.14 	&1.76 	&1.79    &-2244.99602  &1.95 	&1.56 	&2.79  \\
$7.5\times10^{-5}$   &-2244.99090   &2.32 	&1.88 	&1.98    &-2244.99634  &2.02 	&1.61 	&2.86  \\
$5.0\times10^{-5}$   &-2244.99233   &2.16 	&1.76 	&2.17    &-2244.99668  &1.95 	&1.56 	&2.91  \\
$2.5\times10^{-5}$   &-2244.99420   &2.06 	&1.65 	&2.45    &-2244.99706  &1.89 	&1.51 	&2.97  \\
Extrapolated         &        -     &  -    &    -  & -      &-2244.99792  &1.81 	&1.44 	&3.17  \\
&&&&&&&&\\
&&&&\multicolumn{2}{c}{CAS(24,28)}&&&\\
$1.0\times10^{-4}$   &-2244.99504   &2.57 	&2.08 	&1.85     &-2245.00255   &2.36 	&1.86 	&2.99   \\
$7.5\times10^{-5}$   &-2244.99656   &2.78 	&2.27 	&2.09     &-2245.00297   &2.45 	&1.94 	&3.07   \\
$5.0\times10^{-5}$   &-2244.99827   &2.62 	&2.15 	&2.32     &-2245.00343   &2.37 	&1.89 	&3.15   \\
$2.5\times10^{-5}$   &-2245.00052   &2.50 	&2.02 	&2.63     &-2245.00395   &2.31 	&1.83 	&3.22   \\
Extrapolated         &        -     &  -    &    -  &  -      &-2245.00516   &2.21 	&1.75 	&3.47   \\
&&&&&&&&\\
&&&&\multicolumn{2}{c}{CAS(32,35)}&&&\\
$1.0\times10^{-4}$   &-2245.07673  &3.17 	&2.83 	&1.07  &-2245.09709  &2.65 	&2.21 	&2.73  \\
$7.5\times10^{-5}$   &-2245.08070  &3.14 	&2.78 	&1.57  &-2245.09877  &2.66 	&2.20 	&3.00  \\
$5.0\times10^{-5}$   &-2245.08519  &2.86 	&2.57 	&1.86  &-2245.10065  &2.49 	&2.03 	&3.15  \\
$2.5\times10^{-5}$   &-2245.09145  &2.63 	&2.29 	&2.24  &-2245.10316  &2.28 	&1.80 	&3.30  \\
Extrapolated         &       -    & -     & -     & -   &-2245.11142   &1.71 	&1.02 	&4.26  \\
&&&&&&&&\\
&&&&\multicolumn{2}{c}{CAS(40,42)}&&&\\
$1.0\times10^{-4}$   &-2245.15477 &4.21 	&3.94 	&0.31 &-2245.19185 &3.11 	&2.78 	&2.38 \\
$7.5\times10^{-5}$   &-2245.16086 &3.60 	&3.29 	&0.52 &-2245.19468 &3.05 	&2.67 	&2.48 \\
$5.0\times10^{-5}$   &-2245.16882 &3.48 	&3.14 	&0.91 &-2245.19838 &2.88 	&2.50 	&2.67 \\
$2.5\times10^{-5}$   &-2245.18018 &3.29 	&2.95 	&1.37 &-2245.20369 &2.79 	&2.38 	&2.90 \\
Extrapolated         &     -      & -     & -     &  -   &-2245.22419 &2.57 	&1.99 	&4.34 \\\toprule 
			\end{tabular}
		}
		\begin{tablenotes}
			\item[a] See Table \ref{Hexacene} for additional explanations.
		\end{tablenotes}
	\end{threeparttable}\label{FePCAS}
\end{table}

\section{Conclusions and Outlook}\label{Conclusion}
A nearly exact CASSCF approach, iCISCF, has been proposed to treat strongly correlated systems that are beyond the capability of CASSCF.
It combines the selected iCI for the active space solver and the hybrid JR+QN algorithm for orbital optimization. Upon convergence, an inner-space
second-order perturbation step (i.e., iCISCF(2)) can further be taken to improve the energy estimate. The current implementation of iCISCF/iCISCF(2)
can handle ca. 50 active orbitals in conjunction with more than 1000 basis functions, as shown by the showcases of
polyacenes and Fe(II)-porphyrin. Further developments of iCISCF will include (1) the use of $\mathbbm{i}$CAS\cite{iCAS} for automatic construction
and localization of CAS, (2) extrapolation/interpolation schemes for speedup of orbital optimization,
 (3) analytic energy gradients (which are operationally precisely the same as those of CASSCF), and (4) stochastic treatment of
dynamic correlation from the complementary space $Q$.

\section*{Acknowledgement}
This work was supported by the National Natural Science Foundation of China (Grant Nos. 21833001 and 21973054),
Mountain Tai Climb Program of Shandong Province, and Key-Area Research and Development Program of Guangdong Province (Grant No. 2020B0101350001).
YG was further supported by the Qilu Young Scholar Program of Shandong University.

\section*{Data Availability Statement}
The data that supports the findings of this study are available within the article.

\appendix
\renewcommand{\theequation}{A.\arabic{equation}}

\setcounter{equation}{0}

\section{Orbital gradients and diagonal Hessian}\label{GHmat}
By virtue of the basic commutator
\begin{eqnarray}
[\hat{E}_{pq}, \hat{E}_{rs}]=\delta_{qr}\hat{E}_{ps}-\delta_{ps}\hat{E}_{rq}, \label{Commutator}
\end{eqnarray}
the following commutators can be derived
\begin{eqnarray}
[\hat{E}_{pq},\hat{H}]&=&(1-P_{pq}c.c.)\sum_{\sigma}a_{p\sigma}^\dag[a_{q\sigma}, H]\\
&=&\{\hat{E}_{pr}h_{rq}+\hat{e}_{pr,st}g_{rq,st}\}-\{h_{pr}\hat{E}_{rq}+g_{pr,st}\hat{e}_{rq,st}\},\\
\protect[\hat{E}_{pq}^-,\hat{H}\protect]&=&\protect[\hat{E}_{pq},\hat{H}\protect]+c.c.\\
&=&\{(\hat{E}_{pr}+\hat{E}_{rp})h_{rq}+(\hat{e}_{pr,st}+\hat{e}_{rp,st})g_{rq,st}\}\nonumber\\
&-&\{h_{pr}(\hat{E}_{rq}+\hat{E}_{qr})+g_{pr,st}(\hat{e}_{rq,st}+\hat{e}_{qr,st})\},
\end{eqnarray}
in terms of which the orbital gradients \eqref{Orbgrd}, the $\hat{H}_{\kappa}$ Hamiltonian \eqref{Hkoper}, and the sigma vectors \eqref{OrbOrbsigma} and \eqref{OrbCIsigma}
can readily be obtained. Further in view of the following identities
\begin{eqnarray}
D_{ip}&=&2\delta_{ip},\\
\Gamma_{pq,ij}&=&2\delta_{ij}D_{pq}-\delta_{iq}D_{pj},\\
\Gamma_{pi,jq}&=&4\delta_{ip}\delta_{jq}-\delta_{ij}D_{pq},
\end{eqnarray}
the explicit expressions for the orbital gradients \eqref{Orbgrd} and the diagonal elements of the orbital Hessian \eqref{Orbhessian} (which are required by the BFGS algorithm)
can readily be derived
\begin{eqnarray}
G^{\mathrm{o}}_{ia} &=&4f_{ia}^{\mathrm{c}}+4f_{ia}^{\mathrm{a}},\\
G^{\mathrm{o}}_{it} &=&4f_{it}^{\mathrm{c}}+4f_{it}^{\mathrm{a}}-2F_{ti},\\
G^{\mathrm{o}}_{ta} &=&2F_{ta},\\
G^{\mathrm{o}}_{ut} &=&2(F_{ut}-F_{tu}),\\
E^{\mathrm{oo}}_{ia,ia}&=&4\{f_{aa}^{\mathrm{c}}+f_{aa}^{\mathrm{a}}-f_{ii}^{\mathrm{c}}-f_{ii}^{\mathrm{a}}+3K^{ii}_{aa}-J^{ii}_{aa} \},\\
E^{\mathrm{oo}}_{it,it}&=&4\{f_{tt}^{\mathrm{c}}+f_{tt}^{\mathrm{a}}-f_{ii}^{\mathrm{c}}-f_{ii}^{\mathrm{a}}+\frac{1}{2} f_{ii}^{\mathrm{c}}D_{tt}-\frac{1}{2} F_{tt} \nonumber\\
           &+&\sum_u (\delta_{tu}-D_{tu})(3 K^{ii}_{tu} -J^{ii}_{tu})+\frac{1}{2}\sum_{u,v}J^{uv}_{ii}P^{vu}_{tt}
          +\sum_{u,v}K^{uv}_{ii} Q^{vu}_{tt}  \},\\
E^{\mathrm{oo}}_{ta,ta}&=&4\{\frac{1}{2} D_{tt} f_{aa}^{\mathrm{c}}+\frac{1}{2}\sum_{u,v}P^{uv}_{tt}J^{uv}_{aa}+\sum_{u,v}Q^{uv}_{tt}K^{vu}_{aa}
           -\frac{1}{2} F_{tt} \},\\
E^{\mathrm{oo}}_{tu,tu}&=&2(1+P_{tu})W_{tu,tu}-4W_{tu,ut}-2(F_{tt}+F_{uu}),\quad \forall t> u,
\end{eqnarray}
where
\begin{eqnarray}
f_{pq}^{\mathrm{c}}&=&h_{pq}+\sum_{i}(2J^{ii}_{pq} -K^{ii}_{pq}),\\
f_{pq}^{\mathrm{a}}&=&\sum_{u,t}(J^{tu}_{pq} -\frac{1}{2}K^{tu}_{pq})D_{ut},\\
F_{up}&=&\sum_v D_{uv}f^{\mathrm{c}}_{vp}+\sum_{vwt}P^{wt}_{uv}J^{tw}_{vp},\\
W_{tu,tu}&=&D_{tt}f_{uu}^{\mathrm{c}}+\sum_{rs}P^{rs}_{tt}J^{sr}_{uu} +2\sum_{rs}Q^{rs}_{tt}K^{sr}_{uu},\\
W_{tu,ut}&=&D_{tu}f_{ut}^{\mathrm{c}}+\sum_{rs}P^{rs}_{tu}J^{sr}_{ut} +2\sum_{rs}Q^{rs}_{tu}K^{st}_{ut},\\
Q^{vw}_{tu}&=&\frac{1}{2}(\Gamma_{tv,uw}+\Gamma_{vt,uw})=Q^{wv}_{ut},\\
K^{rs}_{pq}&=& g_{pr,sq}=K^{sr}_{qp}.
\end{eqnarray}

\bibliography{iCISCF}

\providecommand{\latin}[1]{#1}
\makeatletter
\providecommand{\doi}
  {\begingroup\let\do\@makeother\dospecials
  \catcode`\{=1 \catcode`\}=2 \doi@aux}
\providecommand{\doi@aux}[1]{\endgroup\texttt{#1}}
\makeatother
\providecommand*\mcitethebibliography{\thebibliography}
\csname @ifundefined\endcsname{endmcitethebibliography}
  {\let\endmcitethebibliography\endthebibliography}{}
\begin{mcitethebibliography}{59}
\providecommand*\natexlab[1]{#1}
\providecommand*\mciteSetBstSublistMode[1]{}
\providecommand*\mciteSetBstMaxWidthForm[2]{}
\providecommand*\mciteBstWouldAddEndPuncttrue
  {\def\EndOfBibitem{\unskip.}}
\providecommand*\mciteBstWouldAddEndPunctfalse
  {\let\EndOfBibitem\relax}
\providecommand*\mciteSetBstMidEndSepPunct[3]{}
\providecommand*\mciteSetBstSublistLabelBeginEnd[3]{}
\providecommand*\EndOfBibitem{}
\mciteSetBstSublistMode{f}
\mciteSetBstMaxWidthForm{subitem}{(\alph{mcitesubitemcount})}
\mciteSetBstSublistLabelBeginEnd
  {\mcitemaxwidthsubitemform\space}
  {\relax}
  {\relax}

\bibitem[Cheung \latin{et~al.}(1978)Cheung, Sundberg, and Ruedenberg]{FORS1978}
Cheung,~L.; Sundberg,~K.; Ruedenberg,~K. Dimerization of carbene to ethylene.
  \emph{J. Am. Chem. Soc.} \textbf{1978}, \emph{100}, 8024--8025\relax
\mciteBstWouldAddEndPuncttrue
\mciteSetBstMidEndSepPunct{\mcitedefaultmidpunct}
{\mcitedefaultendpunct}{\mcitedefaultseppunct}\relax
\EndOfBibitem
\bibitem[Roos \latin{et~al.}(1980)Roos, Taylor, and Siegbahn]{CASSCF}
Roos,~B.~O.; Taylor,~P.~R.; Siegbahn,~P. E.~M. A complete active space scf
  method (casscf) using a density-matrix formulated super-ci approach.
  \emph{Chem. Phys.} \textbf{1980}, \emph{48}, 157--173\relax
\mciteBstWouldAddEndPuncttrue
\mciteSetBstMidEndSepPunct{\mcitedefaultmidpunct}
{\mcitedefaultendpunct}{\mcitedefaultseppunct}\relax
\EndOfBibitem
\bibitem[Werner(1987)]{WernerRev1987}
Werner,~H.-J. Matrix-formulated direct multiconfiguration self-consistent field
  and multiconfiguration reference configuration-interaction methods.
  \emph{Adv. Chem. Phys.} \textbf{1987}, \emph{69}, 1--62\relax
\mciteBstWouldAddEndPuncttrue
\mciteSetBstMidEndSepPunct{\mcitedefaultmidpunct}
{\mcitedefaultendpunct}{\mcitedefaultseppunct}\relax
\EndOfBibitem
\bibitem[Roos(1987)]{CASSCFRev1987}
Roos,~B.~O. The complete active space self-consistent field method and its
  applications in electronic structure calculations. \emph{Adv. Chem. Phys.}
  \textbf{1987}, \emph{69}, 399--445\relax
\mciteBstWouldAddEndPuncttrue
\mciteSetBstMidEndSepPunct{\mcitedefaultmidpunct}
{\mcitedefaultendpunct}{\mcitedefaultseppunct}\relax
\EndOfBibitem
\bibitem[Shepard(1987)]{MCSCFRev1987}
Shepard,~R. The multiconfiguration self-consistent field method. \emph{Adv.
  Chem. Phys} \textbf{1987}, \emph{69}, 63--200\relax
\mciteBstWouldAddEndPuncttrue
\mciteSetBstMidEndSepPunct{\mcitedefaultmidpunct}
{\mcitedefaultendpunct}{\mcitedefaultseppunct}\relax
\EndOfBibitem
\bibitem[Schmidt and Gordon(1998)Schmidt, and Gordon]{MCSCFRev1998}
Schmidt,~M.~W.; Gordon,~M.~S. The construction and interpretation of MCSCF
  wavefunctions. \emph{Annu. Rev. Phys. Chem.} \textbf{1998}, \emph{49},
  233--266\relax
\mciteBstWouldAddEndPuncttrue
\mciteSetBstMidEndSepPunct{\mcitedefaultmidpunct}
{\mcitedefaultendpunct}{\mcitedefaultseppunct}\relax
\EndOfBibitem
\bibitem[Szalay \latin{et~al.}(2012)Szalay, Muller, Gidofalvi, Lischka, and
  Shepard]{MCSCFrev2012}
Szalay,~P.~G.; Muller,~T.; Gidofalvi,~G.; Lischka,~H.; Shepard,~R.
  Multiconfiguration self-consistent field and multireference configuration
  interaction methods and applications. \emph{Chem. Rev.} \textbf{2012},
  \emph{112}, 108--181\relax
\mciteBstWouldAddEndPuncttrue
\mciteSetBstMidEndSepPunct{\mcitedefaultmidpunct}
{\mcitedefaultendpunct}{\mcitedefaultseppunct}\relax
\EndOfBibitem
\bibitem[Sun(2016)]{Co-Iter}
Sun,~Q. Co-iterative augmented Hessian method for orbital optimization.
  \emph{arXiv preprint arXiv:1610.08423} \textbf{2016}, \relax
\mciteBstWouldAddEndPunctfalse
\mciteSetBstMidEndSepPunct{\mcitedefaultmidpunct}
{}{\mcitedefaultseppunct}\relax
\EndOfBibitem
\bibitem[Kreplin \latin{et~al.}(2019)Kreplin, Knowles, and
  Werner]{Werner2020-1}
Kreplin,~D.~A.; Knowles,~P.~J.; Werner,~H.-J. Second-order MCSCF optimization
  revisited. I. Improved algorithms for fast and robust second-order CASSCF
  convergence. \emph{J. Chem. Phys.} \textbf{2019}, \emph{150}, 194106\relax
\mciteBstWouldAddEndPuncttrue
\mciteSetBstMidEndSepPunct{\mcitedefaultmidpunct}
{\mcitedefaultendpunct}{\mcitedefaultseppunct}\relax
\EndOfBibitem
\bibitem[Kreplin \latin{et~al.}(2020)Kreplin, Knowles, and
  Werner]{Werner2020-2}
Kreplin,~D.~A.; Knowles,~P.~J.; Werner,~H.-J. MCSCF optimization revisited. II.
  Combined first- and second-order orbital optimization for large molecules.
  \emph{J. Chem. Phys.} \textbf{2020}, \emph{152}, 074102\relax
\mciteBstWouldAddEndPuncttrue
\mciteSetBstMidEndSepPunct{\mcitedefaultmidpunct}
{\mcitedefaultendpunct}{\mcitedefaultseppunct}\relax
\EndOfBibitem
\bibitem[Vogiatzis \latin{et~al.}(2017)Vogiatzis, Ma, Olsen, Gagliardi, and
  de~Jong]{CAS2222}
Vogiatzis,~K.~D.; Ma,~D.; Olsen,~J.; Gagliardi,~L.; de~Jong,~W.~A. Pushing
  configuration-interaction to the limit: Towards massively parallel MCSCF
  calculations. \emph{J. Chem. Phys.} \textbf{2017}, \emph{147}, 184111\relax
\mciteBstWouldAddEndPuncttrue
\mciteSetBstMidEndSepPunct{\mcitedefaultmidpunct}
{\mcitedefaultendpunct}{\mcitedefaultseppunct}\relax
\EndOfBibitem
\bibitem[Zhang \latin{et~al.}(2021)Zhang, Liu, and Hoffmann]{iCIPT2New}
Zhang,~N.; Liu,~W.; Hoffmann,~M.~R. Further Development of iCIPT2 for Strongly
  Correlated Electrons. \emph{J. Chem. Theory Comput.} \textbf{2021},
  \emph{17}, 949--964\relax
\mciteBstWouldAddEndPuncttrue
\mciteSetBstMidEndSepPunct{\mcitedefaultmidpunct}
{\mcitedefaultendpunct}{\mcitedefaultseppunct}\relax
\EndOfBibitem
\bibitem[Gidofalvi and Mazziotti(2008)Gidofalvi, and Mazziotti]{v2RDM2008}
Gidofalvi,~G.; Mazziotti,~D.~A. Active-space two-electron
  reduced-density-matrix method: Complete active-space calculations without
  diagonalization of the N-electron Hamiltonian. \emph{J. Chem. Phys.}
  \textbf{2008}, \emph{129}, 134108\relax
\mciteBstWouldAddEndPuncttrue
\mciteSetBstMidEndSepPunct{\mcitedefaultmidpunct}
{\mcitedefaultendpunct}{\mcitedefaultseppunct}\relax
\EndOfBibitem
\bibitem[Fosso-Tande \latin{et~al.}(2016)Fosso-Tande, Nguyen, Gidofalvi, and
  DePrince]{v2RDM2016}
Fosso-Tande,~J.; Nguyen,~T.-S.; Gidofalvi,~G.; DePrince,~A.~E. Large-Scale
  Variational Two-Electron Reduced-Density-Matrix-Driven Complete Active Space
  Self-Consistent Field Methods. \emph{J. Chem. Theory Comput.} \textbf{2016},
  \emph{12}, 2260--2271\relax
\mciteBstWouldAddEndPuncttrue
\mciteSetBstMidEndSepPunct{\mcitedefaultmidpunct}
{\mcitedefaultendpunct}{\mcitedefaultseppunct}\relax
\EndOfBibitem
\bibitem[Zgid and Nooijen(2008)Zgid, and Nooijen]{DMRGSCF2008a}
Zgid,~D.; Nooijen,~M. The density matrix renormalization group self-consistent
  field method: Orbital optimization with the density matrix renormalization
  group method in the active space. \emph{J. Chem. Phys.} \textbf{2008},
  \emph{128}, 144116\relax
\mciteBstWouldAddEndPuncttrue
\mciteSetBstMidEndSepPunct{\mcitedefaultmidpunct}
{\mcitedefaultendpunct}{\mcitedefaultseppunct}\relax
\EndOfBibitem
\bibitem[Ghosh \latin{et~al.}(2008)Ghosh, Hachmann, Yanai, and
  Chan]{DMRGSCF2008b}
Ghosh,~D.; Hachmann,~J.; Yanai,~T.; Chan,~G. K.-L. Orbital optimization in the
  density matrix renormalization group, with applications to polyenes and
  beta-carotene. \emph{J. Chem. Phys.} \textbf{2008}, \emph{128}, 144117\relax
\mciteBstWouldAddEndPuncttrue
\mciteSetBstMidEndSepPunct{\mcitedefaultmidpunct}
{\mcitedefaultendpunct}{\mcitedefaultseppunct}\relax
\EndOfBibitem
\bibitem[Yanai \latin{et~al.}(2009)Yanai, Kurashige, Ghosh, and
  Chan]{DMRGSCF2009}
Yanai,~T.; Kurashige,~Y.; Ghosh,~D.; Chan,~G. K.-L. Accelerating convergence in
  iterative solution for large-scale complete active space
  self-consistent-field calculations. \emph{Int. J. Quantum Chem.}
  \textbf{2009}, \emph{109}, 2178--2190\relax
\mciteBstWouldAddEndPuncttrue
\mciteSetBstMidEndSepPunct{\mcitedefaultmidpunct}
{\mcitedefaultendpunct}{\mcitedefaultseppunct}\relax
\EndOfBibitem
\bibitem[Ma and Ma(2013)Ma, and Ma]{DMRGSCF2013}
Ma,~Y.; Ma,~H. Assessment of various natural orbitals as the basis of large
  active space density-matrix renormalization group calculations. \emph{J.
  Chem. Phys.} \textbf{2013}, \emph{138}, 224105\relax
\mciteBstWouldAddEndPuncttrue
\mciteSetBstMidEndSepPunct{\mcitedefaultmidpunct}
{\mcitedefaultendpunct}{\mcitedefaultseppunct}\relax
\EndOfBibitem
\bibitem[Wouters \latin{et~al.}(2014)Wouters, Bogaerts, Van Der~Voort,
  Van~Speybroeck, and Van~Neck]{DMRGSCF2014}
Wouters,~S.; Bogaerts,~T.; Van Der~Voort,~P.; Van~Speybroeck,~V.; Van~Neck,~D.
  Communication: DMRG-SCF study of the singlet, triplet, and quintet states of
  oxo-Mn(Salen). \emph{J. Chem. Phys.} \textbf{2014}, \emph{140}, 241103\relax
\mciteBstWouldAddEndPuncttrue
\mciteSetBstMidEndSepPunct{\mcitedefaultmidpunct}
{\mcitedefaultendpunct}{\mcitedefaultseppunct}\relax
\EndOfBibitem
\bibitem[Ma \latin{et~al.}(2017)Ma, Knecht, Keller, and Reiher]{DMRGSCF2017a}
Ma,~Y.; Knecht,~S.; Keller,~S.; Reiher,~M. Second-order self-consistent-field
  density-matrix renormalization group. \emph{J. Chem. Theory Comput.}
  \textbf{2017}, \emph{13}, 2533--2549\relax
\mciteBstWouldAddEndPuncttrue
\mciteSetBstMidEndSepPunct{\mcitedefaultmidpunct}
{\mcitedefaultendpunct}{\mcitedefaultseppunct}\relax
\EndOfBibitem
\bibitem[Sun \latin{et~al.}(2017)Sun, Yang, and Chan]{DMRGSCF2017b}
Sun,~Q.; Yang,~J.; Chan,~G. K.-L. A general second order complete active space
  self-consistent-field solver for large-scale systems. \emph{Chem. Phys.
  Lett.} \textbf{2017}, \emph{683}, 291--299\relax
\mciteBstWouldAddEndPuncttrue
\mciteSetBstMidEndSepPunct{\mcitedefaultmidpunct}
{\mcitedefaultendpunct}{\mcitedefaultseppunct}\relax
\EndOfBibitem
\bibitem[Thomas \latin{et~al.}(2015)Thomas, Sun, Alavi, and
  Booth]{FCIQMCSCF2015}
Thomas,~R.~E.; Sun,~Q.; Alavi,~A.; Booth,~G.~H. Stochastic Multiconfigurational
  Self-Consistent Field Theory. \emph{J. Chem. Theory Comput.} \textbf{2015},
  \emph{11}, 5316--5325\relax
\mciteBstWouldAddEndPuncttrue
\mciteSetBstMidEndSepPunct{\mcitedefaultmidpunct}
{\mcitedefaultendpunct}{\mcitedefaultseppunct}\relax
\EndOfBibitem
\bibitem[Li~Manni \latin{et~al.}(2016)Li~Manni, Smart, and
  Alavi]{FCIQMCSCF2016}
Li~Manni,~G.; Smart,~S.~D.; Alavi,~A. Combining the Complete Active Space
  Self-Consistent Field Method and the Full Configuration Interaction Quantum
  Monte Carlo within a Super-CI Framework, with Application to Challenging
  Metal-Porphyrins. \emph{J. Chem. Theory Comput.} \textbf{2016}, \emph{12},
  1245--1258\relax
\mciteBstWouldAddEndPuncttrue
\mciteSetBstMidEndSepPunct{\mcitedefaultmidpunct}
{\mcitedefaultendpunct}{\mcitedefaultseppunct}\relax
\EndOfBibitem
\bibitem[Smith \latin{et~al.}(2017)Smith, Mussard, Holmes, and
  Sharma]{HBCISCF2017}
Smith,~J. E.~T.; Mussard,~B.; Holmes,~A.~A.; Sharma,~S. Cheap and near exact
  CASSCF with large active spaces. \emph{J. Chem. Theory Comput.}
  \textbf{2017}, \emph{13}, 5468--5478\relax
\mciteBstWouldAddEndPuncttrue
\mciteSetBstMidEndSepPunct{\mcitedefaultmidpunct}
{\mcitedefaultendpunct}{\mcitedefaultseppunct}\relax
\EndOfBibitem
\bibitem[Yao and Umrigar(2021)Yao, and Umrigar]{HBCISCF2021}
Yao,~Y.; Umrigar,~C. Orbital Optimization in Selected Configuration Interaction
  Methods. \emph{arXiv preprint arXiv:2104.02587} \textbf{2021}, \relax
\mciteBstWouldAddEndPunctfalse
\mciteSetBstMidEndSepPunct{\mcitedefaultmidpunct}
{}{\mcitedefaultseppunct}\relax
\EndOfBibitem
\bibitem[Zimmerman and Rask(2019)Zimmerman, and Rask]{iCASSCF2019}
Zimmerman,~P.~M.; Rask,~A.~E. Evaluation of full valence correlation energies
  and gradients. \emph{J. Chem. Phys.} \textbf{2019}, \emph{150}, 244117\relax
\mciteBstWouldAddEndPuncttrue
\mciteSetBstMidEndSepPunct{\mcitedefaultmidpunct}
{\mcitedefaultendpunct}{\mcitedefaultseppunct}\relax
\EndOfBibitem
\bibitem[Levine \latin{et~al.}(2020)Levine, Hait, Tubman, Lehtola, Whaley, and
  Head-Gordon]{ASCISCF}
Levine,~D.~S.; Hait,~D.; Tubman,~N.~M.; Lehtola,~S.; Whaley,~K.~B.;
  Head-Gordon,~M. CASSCF with Extremely Large Active Spaces Using the Adaptive
  Sampling Configuration Interaction Method. \emph{J. Chem. Theory Comput.}
  \textbf{2020}, \emph{16}, 2340--2354\relax
\mciteBstWouldAddEndPuncttrue
\mciteSetBstMidEndSepPunct{\mcitedefaultmidpunct}
{\mcitedefaultendpunct}{\mcitedefaultseppunct}\relax
\EndOfBibitem
\bibitem[Park(2021)]{ASCISCF2}
Park,~J.~W. Second-Order Orbital Optimization with Large Active Spaces Using
  Adaptive Sampling Configuration Interaction (ASCI) and Its Application to
  Molecular Geometry Optimization. \emph{J. Chem. Theory Comput.}
  \textbf{2021}, \emph{17}, 1522--1534\relax
\mciteBstWouldAddEndPuncttrue
\mciteSetBstMidEndSepPunct{\mcitedefaultmidpunct}
{\mcitedefaultendpunct}{\mcitedefaultseppunct}\relax
\EndOfBibitem
\bibitem[Liu and Hoffmann(2016)Liu, and Hoffmann]{iCI}
Liu,~W.; Hoffmann,~M.~R. iCI: Iterative CI toward full CI. \emph{J. Chem.
  Theory Comput.} \textbf{2016}, \emph{12}, 1169--1178, (E) \textbf{2016},
  \emph{12}, 3000\relax
\mciteBstWouldAddEndPuncttrue
\mciteSetBstMidEndSepPunct{\mcitedefaultmidpunct}
{\mcitedefaultendpunct}{\mcitedefaultseppunct}\relax
\EndOfBibitem
\bibitem[Liu and Hoffmann(2014)Liu, and Hoffmann]{SDS}
Liu,~W.; Hoffmann,~M.~R. SDS: the `static-dynamic-static' framework for
  strongly correlated electrons. \emph{Theor. Chem. Acc.} \textbf{2014},
  \emph{133}, 1481\relax
\mciteBstWouldAddEndPuncttrue
\mciteSetBstMidEndSepPunct{\mcitedefaultmidpunct}
{\mcitedefaultendpunct}{\mcitedefaultseppunct}\relax
\EndOfBibitem
\bibitem[Lei \latin{et~al.}(2017)Lei, Liu, and Hoffmann]{SDSPT2}
Lei,~Y.; Liu,~W.; Hoffmann,~M.~R. Further development of SDSPT2 for strongly
  correlated electrons. \emph{Mol. Phys.} \textbf{2017}, \emph{115},
  2696--2707\relax
\mciteBstWouldAddEndPuncttrue
\mciteSetBstMidEndSepPunct{\mcitedefaultmidpunct}
{\mcitedefaultendpunct}{\mcitedefaultseppunct}\relax
\EndOfBibitem
\bibitem[Zhang \latin{et~al.}(2020)Zhang, Liu, and Hoffmann]{iCIPT2}
Zhang,~N.; Liu,~W.; Hoffmann,~M.~R. Iterative Configuration Interaction with
  Selection. \emph{J. Chem. Theory Comput.} \textbf{2020}, \emph{16},
  2296--2316\relax
\mciteBstWouldAddEndPuncttrue
\mciteSetBstMidEndSepPunct{\mcitedefaultmidpunct}
{\mcitedefaultendpunct}{\mcitedefaultseppunct}\relax
\EndOfBibitem
\bibitem[Huang \latin{et~al.}(2017)Huang, Liu, Xiao, and Hoffmann]{iVI}
Huang,~C.; Liu,~W.; Xiao,~Y.; Hoffmann,~M.~R. iVI: An iterative vector
  interaction method for large eigenvalue problems. \emph{J. Comput. Chem.}
  \textbf{2017}, \emph{38}, 2481--2499, (E) \textbf{2018}, \emph{39}, 338\relax
\mciteBstWouldAddEndPuncttrue
\mciteSetBstMidEndSepPunct{\mcitedefaultmidpunct}
{\mcitedefaultendpunct}{\mcitedefaultseppunct}\relax
\EndOfBibitem
\bibitem[Huang and Liu(2019)Huang, and Liu]{iVI-TDDFT}
Huang,~C.; Liu,~W. iVI-TD-DFT: An iterative vector interaction method for
  exterior/interior roots of TD-DFT. \emph{J. Comput. Chem.} \textbf{2019},
  \emph{40}, 1023--1037, (E) \textbf{2018}, \emph{39}, 338\relax
\mciteBstWouldAddEndPuncttrue
\mciteSetBstMidEndSepPunct{\mcitedefaultmidpunct}
{\mcitedefaultendpunct}{\mcitedefaultseppunct}\relax
\EndOfBibitem
\bibitem[Lei \latin{et~al.}()Lei, Suo, and Liu]{iCAS}
Lei,~Y.; Suo,~B.; Liu,~W. $\mathbbm{i}$CAS: Imposed Automatic Selection and
  Localization of Complete Active Spaces. (unpublished).\relax
\mciteBstWouldAddEndPunctfalse
\mciteSetBstMidEndSepPunct{\mcitedefaultmidpunct}
{}{\mcitedefaultseppunct}\relax
\EndOfBibitem
\bibitem[Wang and Liu(2021)Wang, and Liu]{iOI}
Wang,~Z.; Liu,~W. iOI: an Iterative Orbital Interaction Approach for Solving
  the Self-Consistent Field Problem. \emph{arXiv e-prints} \textbf{2021},
  arXiv--2105\relax
\mciteBstWouldAddEndPuncttrue
\mciteSetBstMidEndSepPunct{\mcitedefaultmidpunct}
{\mcitedefaultendpunct}{\mcitedefaultseppunct}\relax
\EndOfBibitem
\bibitem[Helgaker \latin{et~al.}(2014)Helgaker, Jorgensen, and
  Olsen]{ElectronStructure}
Helgaker,~T.; Jorgensen,~P.; Olsen,~J. \emph{Molecular electronic-structure
  theory}; John Wiley \& Sons, 2014\relax
\mciteBstWouldAddEndPuncttrue
\mciteSetBstMidEndSepPunct{\mcitedefaultmidpunct}
{\mcitedefaultendpunct}{\mcitedefaultseppunct}\relax
\EndOfBibitem
\bibitem[Werner and Meyer(1980)Werner, and Meyer]{Werner-HF-MCSCF}
Werner,~H.-J.; Meyer,~W. A quadratically convergent
  multiconfiguration--self-consistent field method with simultaneous
  optimization of orbitals and CI coefficients. \emph{J. Chem. Phys.}
  \textbf{1980}, \emph{73}, 2342--2356\relax
\mciteBstWouldAddEndPuncttrue
\mciteSetBstMidEndSepPunct{\mcitedefaultmidpunct}
{\mcitedefaultendpunct}{\mcitedefaultseppunct}\relax
\EndOfBibitem
\bibitem[Werner and Knowles(1985)Werner, and Knowles]{SOSCF-Molpro}
Werner,~H.; Knowles,~P.~J. A second order multiconfiguration SCF procedure with
  optimum convergence. \emph{J. Chem. Phys.} \textbf{1985}, \emph{82},
  5053--5063\relax
\mciteBstWouldAddEndPuncttrue
\mciteSetBstMidEndSepPunct{\mcitedefaultmidpunct}
{\mcitedefaultendpunct}{\mcitedefaultseppunct}\relax
\EndOfBibitem
\bibitem[Fletcher(2013)]{BFGSorg}
Fletcher,~R. \emph{Practical methods of optimization}; John Wiley \& Sons,
  2013\relax
\mciteBstWouldAddEndPuncttrue
\mciteSetBstMidEndSepPunct{\mcitedefaultmidpunct}
{\mcitedefaultendpunct}{\mcitedefaultseppunct}\relax
\EndOfBibitem
\bibitem[Fischer and Almlof(1992)Fischer, and Almlof]{BFGS}
Fischer,~T.~H.; Almlof,~J. General methods for geometry and wave function
  optimization. \emph{J. Phys. Chem.} \textbf{1992}, \emph{96},
  9768--9774\relax
\mciteBstWouldAddEndPuncttrue
\mciteSetBstMidEndSepPunct{\mcitedefaultmidpunct}
{\mcitedefaultendpunct}{\mcitedefaultseppunct}\relax
\EndOfBibitem
\bibitem[Chaban \latin{et~al.}(1997)Chaban, Schmidt, and Gordon]{Gordon1997}
Chaban,~G.; Schmidt,~M.; Gordon,~M. Approximate second order method for orbital
  optimization of SCF and MCSCF wavefunctions. \emph{Theor. Chem. Acc.}
  \textbf{1997}, \emph{97}, 88--95\relax
\mciteBstWouldAddEndPuncttrue
\mciteSetBstMidEndSepPunct{\mcitedefaultmidpunct}
{\mcitedefaultendpunct}{\mcitedefaultseppunct}\relax
\EndOfBibitem
\bibitem[Ivanic and Ruedenberg(2003)Ivanic, and Ruedenberg]{Jacobi}
Ivanic,~J.; Ruedenberg,~K. A MCSCF method for ground and excited states based
  on full optimizations of successive Jacobi rotations. \emph{J. Comput. Chem.}
  \textbf{2003}, \emph{24}, 1250--1262\relax
\mciteBstWouldAddEndPuncttrue
\mciteSetBstMidEndSepPunct{\mcitedefaultmidpunct}
{\mcitedefaultendpunct}{\mcitedefaultseppunct}\relax
\EndOfBibitem
\bibitem[Eriksen \latin{et~al.}(2020)Eriksen, Anderson, Deustua, Ghanem, Hait,
  Hoffmann, Lee, Levine, Magoulas, Shen, Tubman, Whaley, Xu, Yao, Zhang, Alavi,
  Chan, Head-Gordon, Liu, Piecuch, Sharma, Ten-no, Umrigar, and
  Gauss]{BlindTest}
Eriksen,~J.~J.; Anderson,~T.~A.; Deustua,~J.~E.; Ghanem,~K.; Hait,~D.;
  Hoffmann,~M.~R.; Lee,~S.; Levine,~D.~S.; Magoulas,~I.; Shen,~J.;
  Tubman,~N.~M.; Whaley,~K.~B.; Xu,~E.; Yao,~Y.; Zhang,~N.; Alavi,~A.; Chan,~G.
  K.-L.; Head-Gordon,~M.; Liu,~W.; Piecuch,~P.; Sharma,~S.; Ten-no,~S.~L.;
  Umrigar,~C.~J.; Gauss,~J. The ground state electronic energy of benzene.
  \emph{J. Phys. Chem. Lett.} \textbf{2020}, \emph{11}, 8922--8929\relax
\mciteBstWouldAddEndPuncttrue
\mciteSetBstMidEndSepPunct{\mcitedefaultmidpunct}
{\mcitedefaultendpunct}{\mcitedefaultseppunct}\relax
\EndOfBibitem
\bibitem[Liu \latin{et~al.}(1997)Liu, Hong, Dai, Li, and Dolg]{BDF1}
Liu,~W.; Hong,~G.; Dai,~D.; Li,~L.; Dolg,~M. The {Beijing} 4-component density
  functional theory program package ({BDF}) and its application to {EuO},
  {EuS}, {YbO} and {YbS}. \emph{Theor. Chem. Acc.} \textbf{1997}, \emph{96},
  75--83\relax
\mciteBstWouldAddEndPuncttrue
\mciteSetBstMidEndSepPunct{\mcitedefaultmidpunct}
{\mcitedefaultendpunct}{\mcitedefaultseppunct}\relax
\EndOfBibitem
\bibitem[Liu \latin{et~al.}(2003)Liu, Wang, and Li]{BDF2}
Liu,~W.; Wang,~F.; Li,~L. \emph{J. Theor. Comput. Chem.} \textbf{2003},
  \emph{2}, 257--272\relax
\mciteBstWouldAddEndPuncttrue
\mciteSetBstMidEndSepPunct{\mcitedefaultmidpunct}
{\mcitedefaultendpunct}{\mcitedefaultseppunct}\relax
\EndOfBibitem
\bibitem[Liu \latin{et~al.}(2004)Liu, Wang, and Li]{BDF3}
Liu,~W.; Wang,~F.; Li,~L. In \emph{Recent Advances in Relativistic Molecular
  Theory}; Hirao,~K., Ishikawa,~Y., Eds.; World Scientific: Singapore, 2004; pp
  257--282\relax
\mciteBstWouldAddEndPuncttrue
\mciteSetBstMidEndSepPunct{\mcitedefaultmidpunct}
{\mcitedefaultendpunct}{\mcitedefaultseppunct}\relax
\EndOfBibitem
\bibitem[Liu \latin{et~al.}(2004)Liu, Wang, and Li]{BDFECC}
Liu,~W.; Wang,~F.; Li,~L. In \emph{Encyclopedia of Computational Chemistry};
  von Ragu\'e~Schleyer,~P., Allinger,~N.~L., Clark,~T., Gasteiger,~J.,
  Kollman,~P.~A., Schaefer~III,~H.~F., Eds.; Wiley: Chichester, UK, 2004\relax
\mciteBstWouldAddEndPuncttrue
\mciteSetBstMidEndSepPunct{\mcitedefaultmidpunct}
{\mcitedefaultendpunct}{\mcitedefaultseppunct}\relax
\EndOfBibitem
\bibitem[Zhang \latin{et~al.}(2020)Zhang, Suo, Wang, Zhang, Li, Lei, Zou, Gao,
  Peng, Pu, Xiao, Sun, Wang, Ma, Wang, Guo, and Liu]{BDFrev2020}
Zhang,~Y.; Suo,~B.; Wang,~Z.; Zhang,~N.; Li,~Z.; Lei,~Y.; Zou,~W.; Gao,~J.;
  Peng,~D.; Pu,~Z.; Xiao,~Y.; Sun,~Q.; Wang,~F.; Ma,~Y.; Wang,~X.; Guo,~Y.;
  Liu,~W. BDF: A relativistic electronic structure program package. \emph{J.
  Chem. Phys.} \textbf{2020}, \emph{152}, 064113\relax
\mciteBstWouldAddEndPuncttrue
\mciteSetBstMidEndSepPunct{\mcitedefaultmidpunct}
{\mcitedefaultendpunct}{\mcitedefaultseppunct}\relax
\EndOfBibitem
\bibitem[Dunning(1989)]{cc-pVDZ}
Dunning,~T.~H. Gaussian basis sets for use in correlated molecular
  calculations. I. The atoms boron through neon and hydrogen. \emph{J. Chem.
  Phys.} \textbf{1989}, \emph{90}, 1007--1023\relax
\mciteBstWouldAddEndPuncttrue
\mciteSetBstMidEndSepPunct{\mcitedefaultmidpunct}
{\mcitedefaultendpunct}{\mcitedefaultseppunct}\relax
\EndOfBibitem
\bibitem[Hachmann \latin{et~al.}(2007)Hachmann, Dorando, Avil{\'e}s, and
  Chan]{DMRG_Radical}
Hachmann,~J.; Dorando,~J.~J.; Avil{\'e}s,~M.; Chan,~G. K.-L. The radical
  character of the acenes: a density matrix renormalization group study.
  \emph{J. Chem. Phys.} \textbf{2007}, \emph{127}, 134309\relax
\mciteBstWouldAddEndPuncttrue
\mciteSetBstMidEndSepPunct{\mcitedefaultmidpunct}
{\mcitedefaultendpunct}{\mcitedefaultseppunct}\relax
\EndOfBibitem
\bibitem[Schriber \latin{et~al.}(2018)Schriber, Hannon, Li, and
  Evangelista]{ACI2018}
Schriber,~J.~B.; Hannon,~K.~P.; Li,~C.; Evangelista,~F.~A. A Combined Selected
  Configuration Interaction and Many-Body Treatment of Static and Dynamical
  Correlation in Oligoacenes. \emph{J. Chem. Theory Comput.} \textbf{2018},
  \emph{14}, 6295--6305\relax
\mciteBstWouldAddEndPuncttrue
\mciteSetBstMidEndSepPunct{\mcitedefaultmidpunct}
{\mcitedefaultendpunct}{\mcitedefaultseppunct}\relax
\EndOfBibitem
\bibitem[Olivares-Amaya \latin{et~al.}(2015)Olivares-Amaya, Hu, Nakatani,
  Sharma, Yang, and Chan]{FeP-DMRG}
Olivares-Amaya,~R.; Hu,~W.; Nakatani,~N.; Sharma,~S.; Yang,~J.; Chan,~G. K.-L.
  The ab-initio density matrix renormalization group in practice. \emph{J.
  Chem. Phys.} \textbf{2015}, \emph{142}, 034102\relax
\mciteBstWouldAddEndPuncttrue
\mciteSetBstMidEndSepPunct{\mcitedefaultmidpunct}
{\mcitedefaultendpunct}{\mcitedefaultseppunct}\relax
\EndOfBibitem
\bibitem[Li~Manni \latin{et~al.}(2016)Li~Manni, Smart, and Alavi]{FeP-Alavi1}
Li~Manni,~G.; Smart,~S.~D.; Alavi,~A. Combining the Complete Active Space
  Self-Consistent Field Method and the Full Configuration Interaction Quantum
  Monte Carlo within a Super-CI Framework, with Application to Challenging
  Metal-Porphyrins. \emph{J. Chem. Theory Comput.} \textbf{2016}, \emph{12},
  1245--1258\relax
\mciteBstWouldAddEndPuncttrue
\mciteSetBstMidEndSepPunct{\mcitedefaultmidpunct}
{\mcitedefaultendpunct}{\mcitedefaultseppunct}\relax
\EndOfBibitem
\bibitem[Li~Manni and Alavi(2018)Li~Manni, and Alavi]{FeP-Alavi2}
Li~Manni,~G.; Alavi,~A. Understanding the Mechanism Stabilizing Intermediate
  Spin States in Fe(II)-Porphyrin. \emph{J. Phys. Chem. A} \textbf{2018},
  \emph{122}, 4935--4947\relax
\mciteBstWouldAddEndPuncttrue
\mciteSetBstMidEndSepPunct{\mcitedefaultmidpunct}
{\mcitedefaultendpunct}{\mcitedefaultseppunct}\relax
\EndOfBibitem
\bibitem[Groenhof \latin{et~al.}(2005)Groenhof, Swart, Ehlers, and
  Lammertsma]{FeP-Geom}
Groenhof,~A.~R.; Swart,~M.; Ehlers,~A.~W.; Lammertsma,~K. Electronic Ground
  States of Iron Porphyrin and of the First Species in the Catalytic Reaction
  Cycle of Cytochrome P450s. \emph{J. Phys. Chem. A} \textbf{2005}, \emph{109},
  3411--3417\relax
\mciteBstWouldAddEndPuncttrue
\mciteSetBstMidEndSepPunct{\mcitedefaultmidpunct}
{\mcitedefaultendpunct}{\mcitedefaultseppunct}\relax
\EndOfBibitem
\bibitem[Pierloot(2003)]{FeP-PIERLOOT1}
Pierloot,~K. The CASPT2 method in inorganic electronic spectroscopy: from ionic
  transition metal to covalent actinide complexes. \emph{Mol. Phys.}
  \textbf{2003}, \emph{101}, 2083--2094\relax
\mciteBstWouldAddEndPuncttrue
\mciteSetBstMidEndSepPunct{\mcitedefaultmidpunct}
{\mcitedefaultendpunct}{\mcitedefaultseppunct}\relax
\EndOfBibitem
\bibitem[Weser \latin{et~al.}(2021)Weser, Freitag, Guther, Alavi, and
  Li~Manni]{FeP-Alavi3}
Weser,~O.; Freitag,~L.; Guther,~K.; Alavi,~A.; Li~Manni,~G. Chemical insights
  into the electronic structure of Fe(II) porphyrin using FCIQMC, DMRG, and
  generalized active spaces. \emph{Int. J. Quantum Chem.} \textbf{2021},
  \emph{121}, e26454\relax
\mciteBstWouldAddEndPuncttrue
\mciteSetBstMidEndSepPunct{\mcitedefaultmidpunct}
{\mcitedefaultendpunct}{\mcitedefaultseppunct}\relax
\EndOfBibitem
\end{mcitethebibliography}

\newpage
For TOC only

\includegraphics[width=\textwidth]{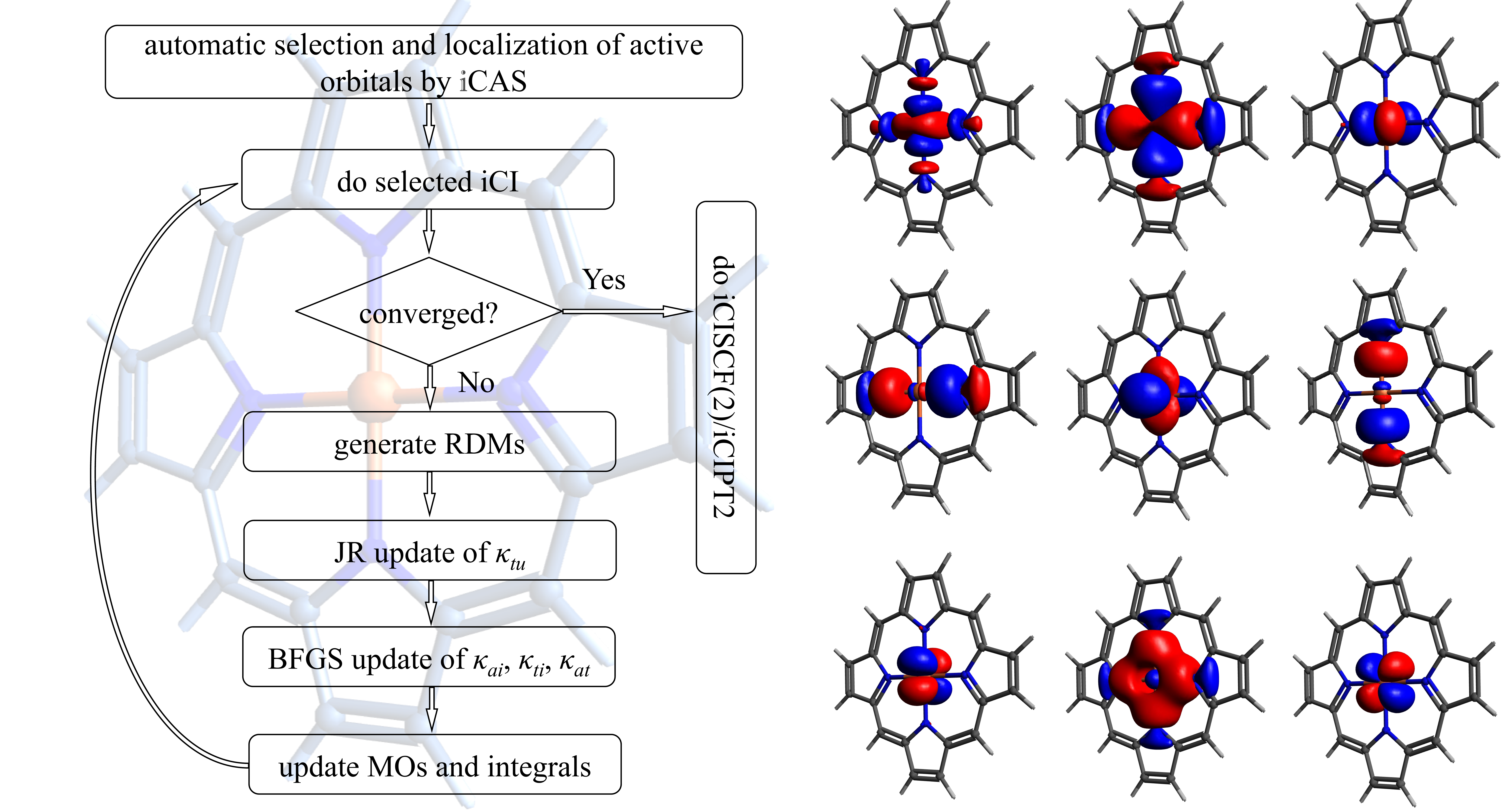}

\end{document}